\documentclass[twocolumn,tighten]{aastex6}
\usepackage{privon_astro}

\newcommand{\source}{IRAS~13120--5453\xspace}

\slugcomment{Accepted for publication in the Astrophysical Journal. Draft: \today}

\shorttitle{Dense Molecular Gas in \source}
\shortauthors{Privon et al.}

\begin{document}

\title{The Dense Molecular Gas and Nuclear Activity in the ULIRG \source}
\author{G.~C.~Privon\altaffilmark{1,2}}
\author{S.~Aalto\altaffilmark{3}}
\author{N.~Falstad\altaffilmark{3}}
\author{S.~Muller\altaffilmark{3}}
\author{E.~Gonz\'alez-Alfonso\altaffilmark{4}}
\author{K.~Sliwa\altaffilmark{5}}
\author{E.~Treister\altaffilmark{1,2}}
\author{F.~Costagliola\altaffilmark{3}}
\author{L.~Armus\altaffilmark{6}}
\author{A.~S.~Evans\altaffilmark{7,8}}
\author{S.~Garcia-Burillo\altaffilmark{9}}
\author{T.~Izumi\altaffilmark{10}}
\author{K.~Sakamoto\altaffilmark{11}}
\author{P.~van~der~Werf\altaffilmark{12}}
\author{J.~K.~Chu\altaffilmark{13}}

\altaffiltext{1}{Instituto de Astrof\'sica, Facultad de F\'isica, Pontificia Universidad Cat\'olica de Chile, Casilla 306, Santiago 22, Chile}
\altaffiltext{2}{Departamento de Astronom\'ia, Universidad de Concepci\'on, Casilla 160-C, Concepci\'on, Chile}
\altaffiltext{3}{Department of Earth and Space Sciences, Chalmers University of Technology, Onsala Space Observatory, S-439 94 Onsala, Sweden}
\altaffiltext{4}{Universidad de Alcal\'a, Departamento de F\'isica y Matem\'aticas, Campus Universitario, E-28871 Alcal\'a de Henares, Madrid, Spain}
\altaffiltext{5}{Max Planck Institute for Astronomy, K\"onigstuhl 17, D-69117 Heidelberg, Germany}
\altaffiltext{6}{Spitzer Science Center, California Institute of Technology, MS 220-6, Pasadena, CA, 91125, USA}
\altaffiltext{7}{Department of Astronomy, University of Virginia, Charlottesville, VA 22903, USA}
\altaffiltext{8}{National Radio Astronomy Observatory, Charlottesville, VA, 22903 USA}
\altaffiltext{9}{Observatorio de Madrid, OAN-IGN, Alfonso XII, 3, E-28014-Madrid, Spain}
\altaffiltext{10}{Institute of Astronomy, School of Science, The University of Tokyo, 2-21-1 Osawa, Mitaka, Tokyo 181-0015, Japan}
\altaffiltext{11}{Institute of Astronomy and Astrophysics, Academia Sinica, PO Box 23-141, 10617, Taipei, Taiwan}
\altaffiltext{12}{Leiden Observatory, Leiden University, PO Box 9513, 2300 RA Leiden, The Netherlands}
\altaffiltext{13}{Institute for Astronomy, University of Hawaii, 2680 Woodlawn Drive, Honolulu, HI 96822}

\begin{abstract}
We present new ALMA Band 7 ($\sim340$ GHz) observations of the dense gas tracers HCN, HCO$^+$, and CS in the local, single-nucleus, ultraluminous infrared galaxy \source.
We find centrally enhanced \HCN{4}{3} emission, relative to \HCO{4}{3}, but do not find evidence for radiative pumping of HCN.
Considering the size of the starburst (0.5 kpc) and the estimated supernovae rate of $\sim1.2$ \pyr, the high HCN/HCO$^+$ ratio can be explained by an enhanced HCN abundance as a result of mechanical heating by the supernovae, though the active galactic nucleus and winds may also contribute additional mechanical heating.
The starburst size implies a high $\Sigma_{IR}$ of $4.7\times10^{12}$ \Lsun kpc$^{-2}$, slightly below predictions of radiation-pressure limited starbursts.
The HCN line profile has low-level wings, which we tentatively interpret as evidence for outflowing dense molecular gas.
However, the dense molecular outflow seen in the HCN line wings is unlikely to escape the galaxy and is destined to return to the nucleus and fuel future star formation.
We also present modeling of Herschel observations of the \hho lines and find a nuclear dust temperature of $\sim40$ K.
\source has a lower dust temperature and $\Sigma_{IR}$ than is inferred for the systems termed ``compact obscured nuclei'' (such as Arp~220 and Mrk~231).
If \source has undergone a compact obscured nucleus phase, we are likely witnessing it at a time when the feedback has already inflated the nuclear ISM and diluted star formation in the starburst/AGN core.
\end{abstract}


\section{Overview of \source and Dense Gas Tracers}
\label{sec:Introduction}

Star formation rates on $\sim1$ kpc scales are well correlated with the local (molecular) gas surface density \citep{Bigiel2008}, consistent with a scenario in which the stars form out of molecular gas.
As cold \HH does not have strong emission, tracer molecules such as CO, HCN, and HCO$^+$ are used to quantify the properties of the molecular ISM.
Molecular outflows have been identified in many starbursts and active galactic nuclei \citep[AGN; e.g.,][]{Feruglio2010,Alatalo2011,Sturm2011,Aalto2012,Aalto2012a,Veilleux2013,Feruglio2013b,Sakamoto2014,Cicone2014,GarciaBurillo2015} and may represent the clearing of the fuel for star formation.

The present study is focused on the properties of the high-density tracers HCN and HCO$^+$, \hho emission, and the excitation and kinematics of those tracers in the ultraluminous infrared galaxy (ULIRG), \source.
Below we discuss these tracers and the general properties of this ULIRG.

\subsection{HCN, HCO$^+$, and the Star-forming Molecular Gas}
\label{sec:moloverview}

\CO{1}{0} is widely used as a tracer of the total molecular gas mass within a galaxy \citep[e.g.,][]{Bolatto2013}; its relatively low critical density ($n_{crit}\approx10^2$ \cmc) and energy level ($E/k_B=5.5$ K) means it is associated with even the low density molecular gas which is not directly involved in ongoing star formation.
Empirically, this has been seen in studies which show the CO luminosity has a non-linear relation with the star formation rate \citep[as traced by \LIRr;][]{Gao2004a}.
In contrast, the \HCN{1}{0} and \HCO{1}{0} emission are linearly correlated with the SFR \citep[e.g.,][]{Solomon1992a,Gao2004a}.
This, plus the comparatively higher critical densities of the $1\rightarrow0$ lines ($n_{crit}\approx10^6$ and $10^5$ \cmc at 30 K, respectively) suggests the HCN and HCO$^+$ emission trace the dense gas which is actively associated with ongoing star formation.

The excitation of HCN and HCO$^+$ is uncertain in extreme star-forming galaxies.
The excitation seems to systematically vary with gas density and the incident UV radiation field \citep{Meijerink2007} and may vary with the infrared radiation field due to radiative pumping \citep{Aalto1995}.
Additionally, the relative HCN/HCO$^+$ abundance can be affected by chemistry driven by X-rays \citep[e.g.,][]{Lepp1996} and mechanical heating \citep{Loenen2008,Kazandjian2012}.
Understanding the excitation and abundance of these high $n_{crit}$ tracers is crucial to accurately characterize the dense molecular gas in star forming systems (i.e., determining the dense gas fraction and the physical conditions of the dense gas).

Studies of AGN hosts have found evidence for enhanced HCN emission (relative to HCO$^+$) in both galaxy-integrated and resolved observations \citep[e.g.,][]{Kohno2001,Imanishi2006,Imanishi2007,Davies2012}, which has been interpreted as evidence for the influence of X-ray dominated regions (XDRs) or mechanical heating \citep{Izumi2016}.
More recent studies of galaxy-integrated emission have uncovered enhanced HCN emission in pure starburst and composite systems \citep{Costagliola2011,Privon2015}, but existing data were not sufficient to suggest a single preferred physical process for the enhancement.
Other studies have found evidence for non-linear relationships of \HCN{1}{0} with \LFIRr \citep{GarciaBurillo2012}, in contrast to the \citet{Gao2004a} picture.

Recent interferometric observations of the (3--2) and (4--3) lines in systems with enhanced HCN emission have uncovered convincing evidence of infrared pumping, via the detection of $v_2=1f$ lines of HCN \citep[e.g.,][]{Sakamoto2010,Costagliola2013,Imanishi2013,Aalto2015,Aalto2015a}.
The $v_2=1f$ \HCN{4}{3} line has a level energy of $1050$ K, and is thus unlikely to be collisionally excited.
Instead, it has been proposed that mid-infrared pumping, via absorption of $14~\mu m$ photons, excites this ro-vibrational branch \citep{Ziurys1986,Aalto1995}.
It is possible that the radiative pumping may enhance the $v_2=0$ emission of lower-J transitions \citep[e.g.,][]{Carroll1981}, potentially explaining elevated HCN/HCO$^+$ ratios, but this has not yet been confirmed observationally.

Most of the extragalactic HCN $v_2=1f$ detections are in systems which appear to contain dense, high column, hot ($T_{dust}>100$ K) cores \citep{Aalto2015a}.
These systems, dubbed ``Compact Obscured Nuclei'' (CONs), feature compact starbursts and perhaps also deeply buried (Compton-thick) AGN and appear to be optically thick in the mid-infrared.
As a result,  sub-millimeter lines may be the only way to probe the inner structure.

Individual galactic star forming regions have also been found to have elevated HCN/HCO$^+$, particularly in the circumnuclear disk (CND) of the Galactic Center where HCN/HCO$^+\sim1.5-2$ \citep{Mills2013}.
Despite the presence of $v_2=1f$ emission in the Galaxy's CND, enhanced HCN emission appears to not be driven by the IR pumping or XDRs.
\citet{Mills2013} state PDR models are consistent with the observed ratio, but would likely fail to explain the high gas temperatures seen in CND \citep{Requena-Torres2012}, while mechanical heating can simultaneously explain both the HCN/HCO$^+$ ratio and the gas temperature.
Furthermore, mechanical heating from a jet or outflow has been invoked as the driver of high HCN/HCO$^+$ ratios in AGN hosts \citep[e.g.,][]{Izumi2013,Garcia-Burillo2014,Izumi2015,Izumi2016}.

These observational results point to a complex interplay of excitation effects and chemistry-driven abundance variations.
The shocks and turbulence resulting from supernovae as well as AGN and starburst-driven winds, which can penetrate deep into molecular clouds, may result in elevated HCN/HCO$^+$ abundance ratios and higher HCN/HCO$^+$ luminosity ratios.
Substantial variations in the relative abundances of these tracer molecules and their excitation would bias estimates of the dense gas mass from HCN luminosities.
ALMA observations are needed to resolve the emission from these molecular tracers and link luminosity variations to the underlying nuclear star formation and AGN activity.

\subsection{Water Emission}

Emission from \hho molecules appears to be common in extreme star forming galaxies \citep{Fischer1999,Gonzalez-Alfonso2004,Gonzalez-Alfonso2008,Fischer2010,Gonzalez-Alfonso2012}.
The sub-millimeter \hho lines appear to require pumping from the far-infrared continuum \citep{Gonzalez-Alfonso2014}, making water emission a good probe of warm, dusty regions and the far-infrared radiation field in those regions.
Based on modeling of the \hho lines and agreement with results from millimeter HCN observations, \citet{Gonzalez-Alfonso2014} argue the water emission is co-spatial with HCN emission.
Additionally they find the broad characteristics of the \hho emission at submillimeter wavelengths in warm star forming galaxies can be explained with dust temperatures T$_{dust}=55-75$ K, a $100~\mu m$ optical depth $\tau_{100}\sim0.1$, and a column density of $N_{H_2O}\sim(0.2-2)\times10^{17}$ \cmcol, when the highest-lying submillimeter lines (at $>400$ K) are not detected.

The coupling of \hho emission to the infrared radiation field and the co-spatial nature with the HCN suggests that modeling of the water emission can constrain the dust temperature in the dense molecular gas, independently of the infrared SED.
This provides vital constraints on the physical conditions in the molecular regions traced by HCN, aiding in the interpretation of the HCN emission, both for systems that are optically thin and optically thick at $100~\mu m$ \citep{Gonzalez-Alfonso2014}.

\subsection{Target: \source}

\begin{figure}
\includegraphics[width=0.45\textwidth]{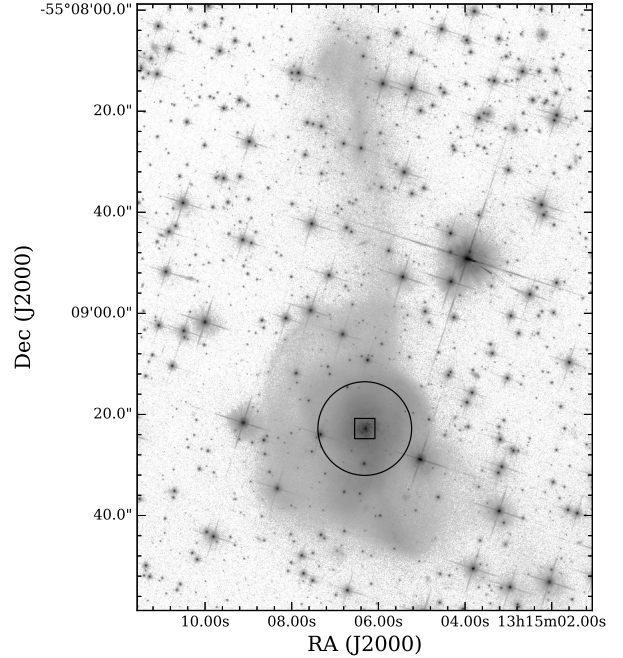}
\caption{HST/ACS F814W image of \source (A. S. Evans et al. \emph{in preparation}), showing the single nucleus, long tidal tail stretching to the north, and the loops from re-accreted tidal material.
The black circle denotes the ALMA Band 7 primary beam.
The detected line and continuum emission is concentrated on the nucleus and confined within the black square shown, which denotes the coverage of the panels in Figures~\ref{fig:maps} and \ref{fig:continuum}.
}
\label{fig:optical}
\end{figure}

\source is a ULIRG with \LIRr $= 2.1\times10^{12}$ \Lsun \citep{Armus2009} at a distance, $D_L=144$ Mpc ($z=0.03112$; angular scale: $0.656$ kpc/arcsec).
Several multiwavelength studies have morphologically classified this system as a post-merger, single nucleus system \citep[Figure~\ref{fig:optical};][]{Haan2011,Stierwalt2013}.
\citet{Kim2013} applied GALFIT modeling to HST images of the system and found a significant portion of the flux ($\sim35\%$) is in non-axisymmetric structures, consistent with a scenario in which the system has not fully relaxed.
A visual inspection of the large-scale morphology shows a faint tidal tail stretching to the north, with multiple loops surrounding the main body of the galaxy, suggesting the extended regions of the system are re-accreting material from the tidal tails.

The system is optically classified as a Seyfert 2 \citep{VeronCetty2001}, and the equivalent width of the $6.2~\mu m$ polycyclic aromatic hydrocarbon \citep[$=0.45~\mu m$;][]{Stierwalt2013} suggests the infrared luminosity arises due to a mix of reprocessed radiation from both a starburst and AGN.
X-ray observations of \source also find evidence for an AGN, with an estimated $\sim18$\% contribution of the AGN to \LIR \citep{Iwasawa2011}.
Nuclear Spectroscopic Telescope Array (NuSTAR) hard X-ray
observations of the system are consistent with the presence of a Compton-thick AGN \citep[$N_H=3.15^{+2.23}_{-1.29}\times10^{24}$ cm$^{-2}$;][]{Teng2015} with L$_{AGN,2-10~keV}=1.25\times10^{43}$ erg s$^{-1}$ and a star formation rate of $\sim170$ \Msun yr$^{-1}$ (the latter determined from the thermal emission and the emission associated with high-mass X-ray binaries).
The high obscuration towards the X-ray emitting region is consistent with the optical classification, where we only see the narrow lines.
The optical depth of the $9.7~\mu m$ silicate absorption feature is $\tau_{9.7}=2.52$ \citep{Stierwalt2014}, corresponding to A$_V\approx23$ following the relationship found by \citet{Roche1985} for $\tau_{9.7}$ and A$_V$ for the galactic center.
We note the obscuration giving rise to the silicate absorption likely occurs outside of the nucleus but within the host galaxy \citep[e.g.,][]{Gonzalez-Martin2013,Roche2015} or from absorption within the starburst \citep{Diaz-Santos2013}.
The \cii emission is suppressed relative to the far-infrared \citep[\cii/\LFIR$=(6.3\pm0.1)\times10^{-4}$;][]{Diaz-Santos2013}.
Spitzer observations were used to place an upper limit on the mid-infrared size of the starburst of 2.68 kpc \citep{Diaz-Santos2010}, leading to a lower limit on the infrared luminosity surface density of $3.0\times10^{11}$ \Lsun pc$^{-2}$.
Using the upper limit on the size, \source lies below the compact starburst / \cii suppression model of \citet{Diaz-Santos2013}, though this may be due to the underestimation of the IR luminosity surface density.

In this paper we present new Atacama Large Millimeter/submillimeter Array (ALMA) Band 7 observations of the $v_2=0,1$ \HCN{4}{3}, $v_2=0$ \HCO{4}{3}, and \CS{7}{6} lines (Section~\ref{sec:obs}), then discuss the excitation of these dense gas tracers and a tentatively detected outflow (Section~\ref{sec:dense}).
Next we show results from modeling of the \hho lines observed with Herschel (Section~\ref{sec:water}), aimed at constraining the dust temperature of the dense ISM.
We then explore the nuclear kinematics of the system (Section~\ref{sec:kinematics}), and the ISM properties and starburst size inferred from our ALMA detection of the $\sim333$ GHz continuum emission (Section~\ref{sec:continuum}).
We conclude by discussing the implications of the ISM properties for the fate of the starburst in \source (Section~\ref{sec:activity}).
Where appropriate, values were computed assuming a WMAP-5 cosmology \citep[H$_0$ = 70 \kms Mpc$^{-1}$, $\Omega_{\textnormal{vacuum}}=0.72$, $\Omega_{\textnormal{matter}}=0.28$;][]{Hinshaw2009}, with corrections for the 3-attractor model of \citet{Mould2000}.

\section{Observations}
\label{sec:obs}

\subsection{Atacama Large Millimeter/submillimeter Array}
\label{sec:ALMA}

ALMA observations were carried out on 18 May 2014 in the C32-5 configuration, as part of project \#2012.1.00817.S (PI: Aalto) with an on-source time of 19.1 minutes.
These data have projected baseline lengths between $23-625$m.
The observing setup consisted of four independent spectral windows: one each tuned to the redshifted frequencies of \HCN{4}{3}, \HCO{4}{3} and two centered at frequencies of 331.9 GHz (covering \CS{7}{6}) and 333.7 GHz (Figure~\ref{fig:profiles}).
All four spectral windows had bandwidths of $1.875$ GHz.
The weather conditions were good, with a precipitable amount of water vapor of 0.8 mm.
The median on-source system temperature was $180$ K.
Observations were calibrated and imaged in a standard fashion using the Common Astronomy Software Applications \citep{McMullin2007}.
The bandpass response of the array was calibrated using the quasar J1037-2934. The flux calibration was set by observations of Ganymede, using the
Butler-JPL-Horizons 2012 model, as described in the ALMA Memo 594 \footnote{\url{https://science.nrao.edu/facilities/alma/aboutALMA/Technology/ALMA_Memo_Series/alma594/abs594}}.
The absolute flux calibration is expected to be better than 10\%.
The gain calibration was done with the quasar J1329-5608.
An iteration of phase self-calibration was possible since the continuum of IRAS13120-5453 is strong enough to allow us to derive gain solutions on a time interval of 20s.
The phase center is 13h15m06.316s --55d09m22.79s (J2000).
The data were imaged using Briggs weighting \citep[robust=0.5;][]{Briggs1995} and the resulting resolution of the data cubes is $0\arcsec.50\times0\arcsec.28$ ($\sim325\times180$ pc) at a position angle of $-75^{\circ}$ with an RMS sensitivity of 1.2 \mJybeam at 20 \kms spectral resolution.
Continuum-free cubes were created by subtracting a linear baseline fit to the line-free channels in the image plane (CASA task \texttt{imcontsub}).
We adopt a rest frequency of $354.526$ GHz for \HCN{4}{3}, $356.754$ GHz for \HCO{4}{3}, and $342.883$ GHz for \CS{7}{6}.
The $v_2=1$ \HCN{4}{3} doublet, included in the two higher frequency spectral windows, have rest frequencies of $354.460$ GHz ($v_2=1e$) and $356.256$ GHz ($v_2=1f$); the former component is underneath the $v_2=0$ line, while the latter is $420$ \kms from the \HCO{4}{3} line (see Figure~\ref{fig:profiles}).
When discussing the vibrational lines, we refer to them as HCN vibrational or explicitly denote the line as $v_2=1f$.
The $v=0$ rotational lines will be referred to with their J-level transitions.
All rest frequencies were obtained from the JPL Submillimeter, Millimeter, and Microwave Spectral Line Catalog \citep{Pickett1998} through Splatalogue.

\subsection{Herschel Space Observatory}
\label{sec:herschel}

\source was observed using the Photodetector Array Camera and Spectrometer \citep[PACS;][]{Poglitsch2010} and the Spectral and Photometric Imaging Receiver \citep[SPIRE;][]{Griffin2010} on the Herschel Space Observatory.
The PACS observations were performed in high spectral sampling range spectroscopy mode on 19 July 2012 as part of the Hermolirg OT2 project (PI: E. Gonz\'alez-Alfonso; ObsIDs: 1342248346 \& 1342248347).

The SPIRE observation was performed on 05 January 2011 as part of the OT key program Hercules (PI: P.P. van der Werf; ObsID: 1342212342) with a single pointing centered on IRAS 13120-5453.
The observation was conducted in high spectral resolution, sparse image sampling mode with a resolution of $1.2$~GHz in both observing bands ($447-989$~GHz and $958-1545$~GHz).
A total of 29 repetitions (58 FTS scans) were performed, resulting in a total on-source integration time of $3863$~s.

The data reduction was done with the Herschel interactive processing environment \citep[HIPE;][]{Ott2010} version 14.0.1.
The PACS observations were reduced using the standard telescope normalization pipeline for chopped line scans and short range scans.
Each spaxel in the PACS $5$~x~$5$ spaxel array is a square with $9.4''$ to each side.
At a distance of $144$~Mpc this corresponds to $\sim6$~kpc, making the nuclear far-IR emission in IRAS 13120-5453 spatially unresolved in the central spaxel.
As the point spread function of the spectrometer is larger than the central spaxel, the central spectrum was extracted using the point source correction task available in HIPE 14.0.1.
To compensate for small pointing offsets and jitter that might move flux out of the central spaxel, this extracted spectrum was scaled to the integrated flux level of the central $3$~x~$3$ spaxels.
The data reduction for the SPIRE observation was done with the standard single pointing pipeline.

To extract the line fluxes of the SPIRE observation a bootstrap method was used.
A total of 58 scans were drawn randomly, with replacement, from the original observation and then averaged together.
For each detector, a polynomial baseline was then subtracted from the spectrum before simultaneously fitting the spectral lines using Gaussian profiles convolved with the instrumental response (a sinc function).
This procedure was repeated 1000 times and a Gaussian was fitted to the resulting flux distribution of each line to obtain its mean line flux and standard deviation.

\section{The Dense Gas Tracers}
\label{sec:dense}

\begin{figure*}
\includegraphics[width=\textwidth]{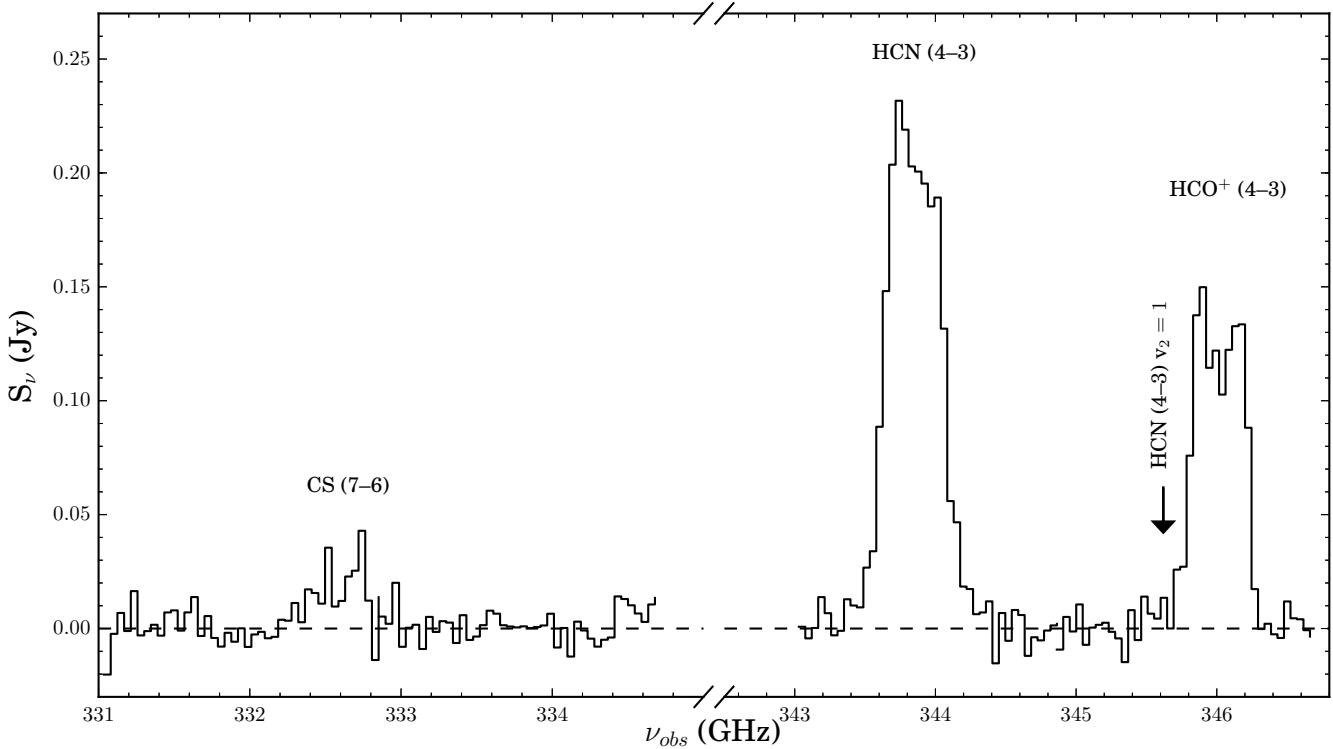}
\caption{The integrated spectra from the ALMA observations, continuum-subtracted and measured from a 3 arcsecond diameter circular region centered on the nucleus.
The locations of detected and expected lines are marked; their measured properties are given in Table~\ref{table:meas}.
The arrow and label marks the location of the \HCN{4}{3} $v_2=1f$ line; emission is seen at those frequencies, but we attribute it to \HCO{4}{3} emission associated with a molecular outflow (Section~\ref{sec:pumping}).}
\label{fig:profiles}
\end{figure*}

\begin{figure*}
\centering
\includegraphics[width=0.47\textwidth]{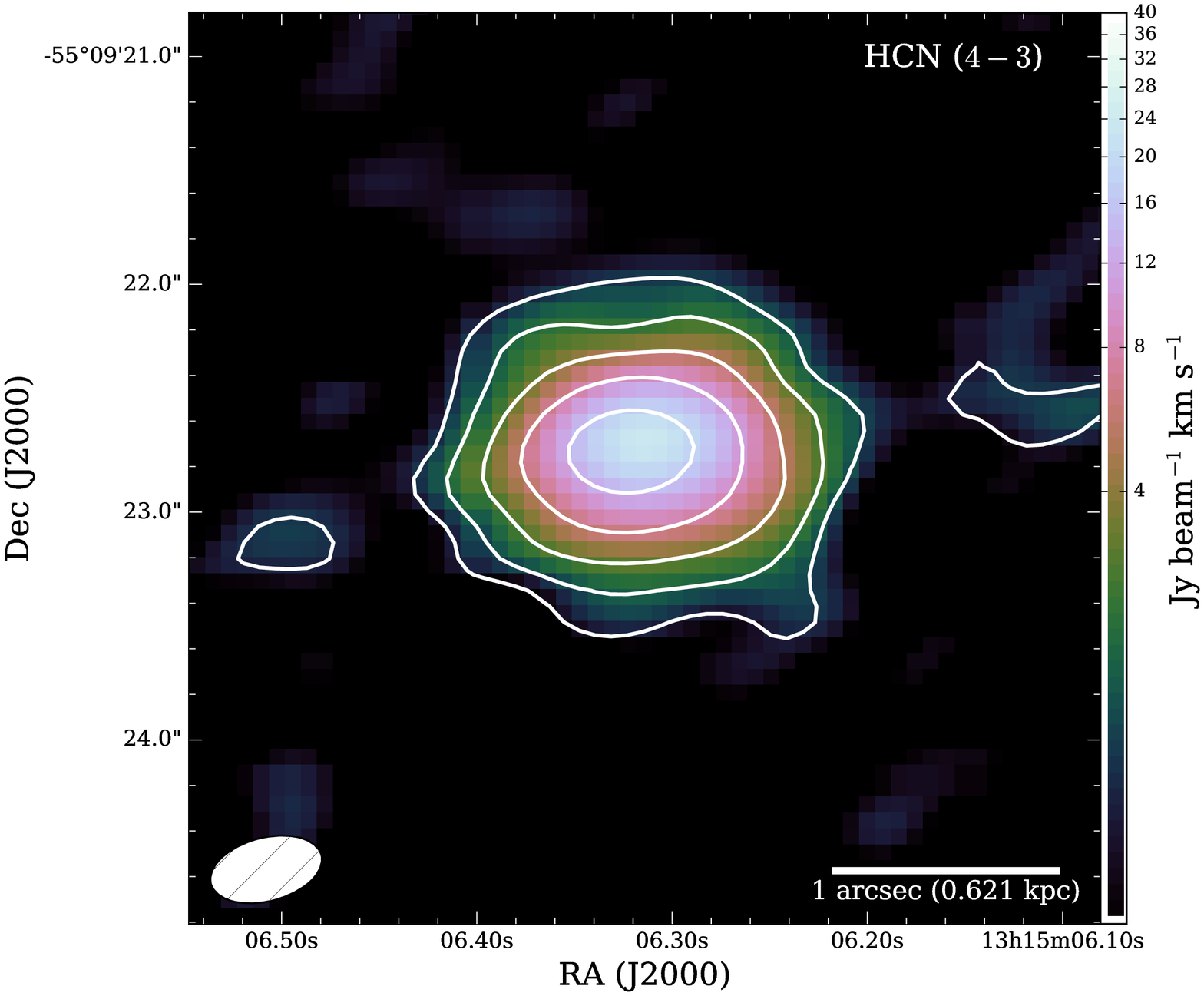}
\includegraphics[width=0.47\textwidth]{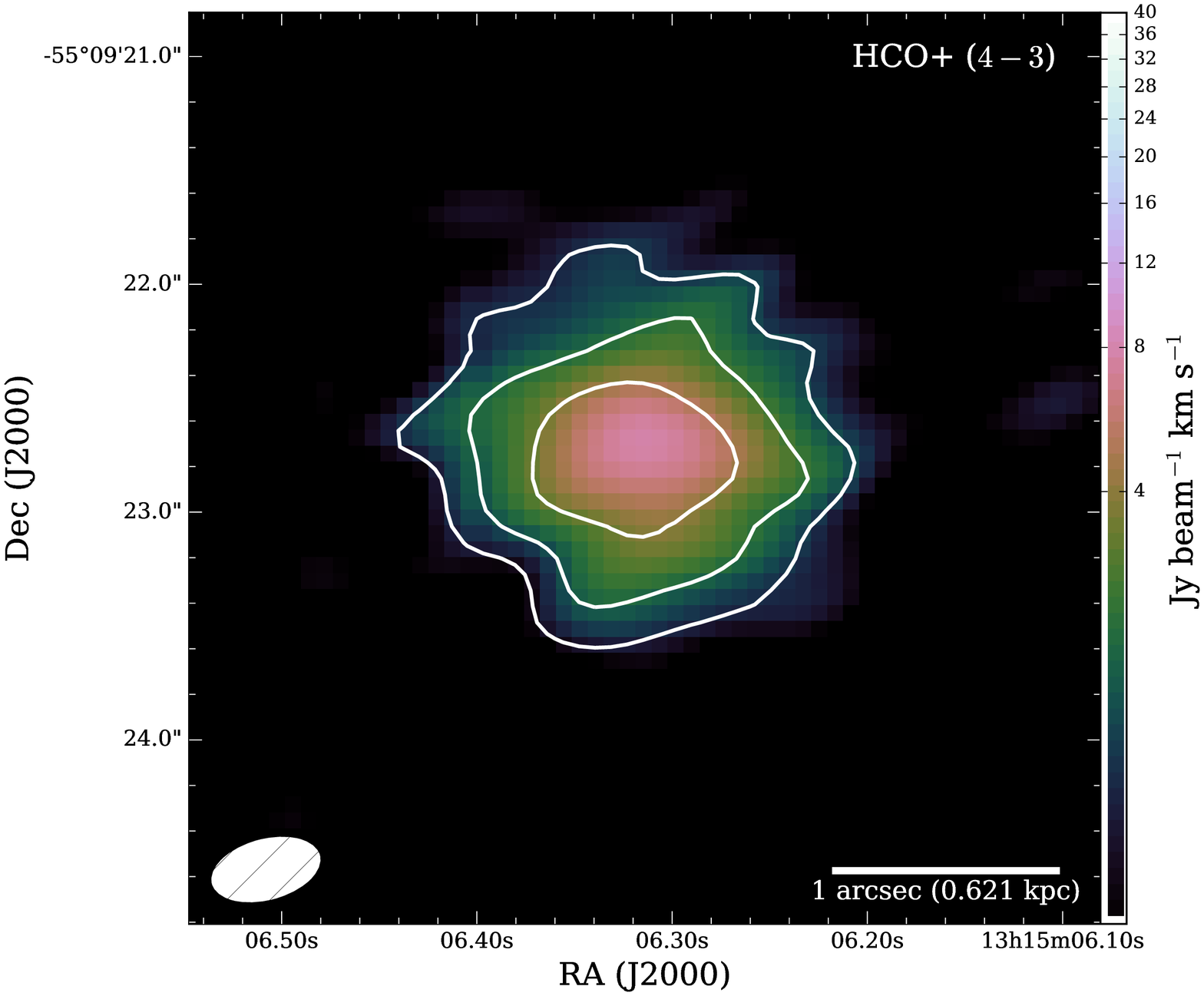}\\
\includegraphics[width=0.47\textwidth]{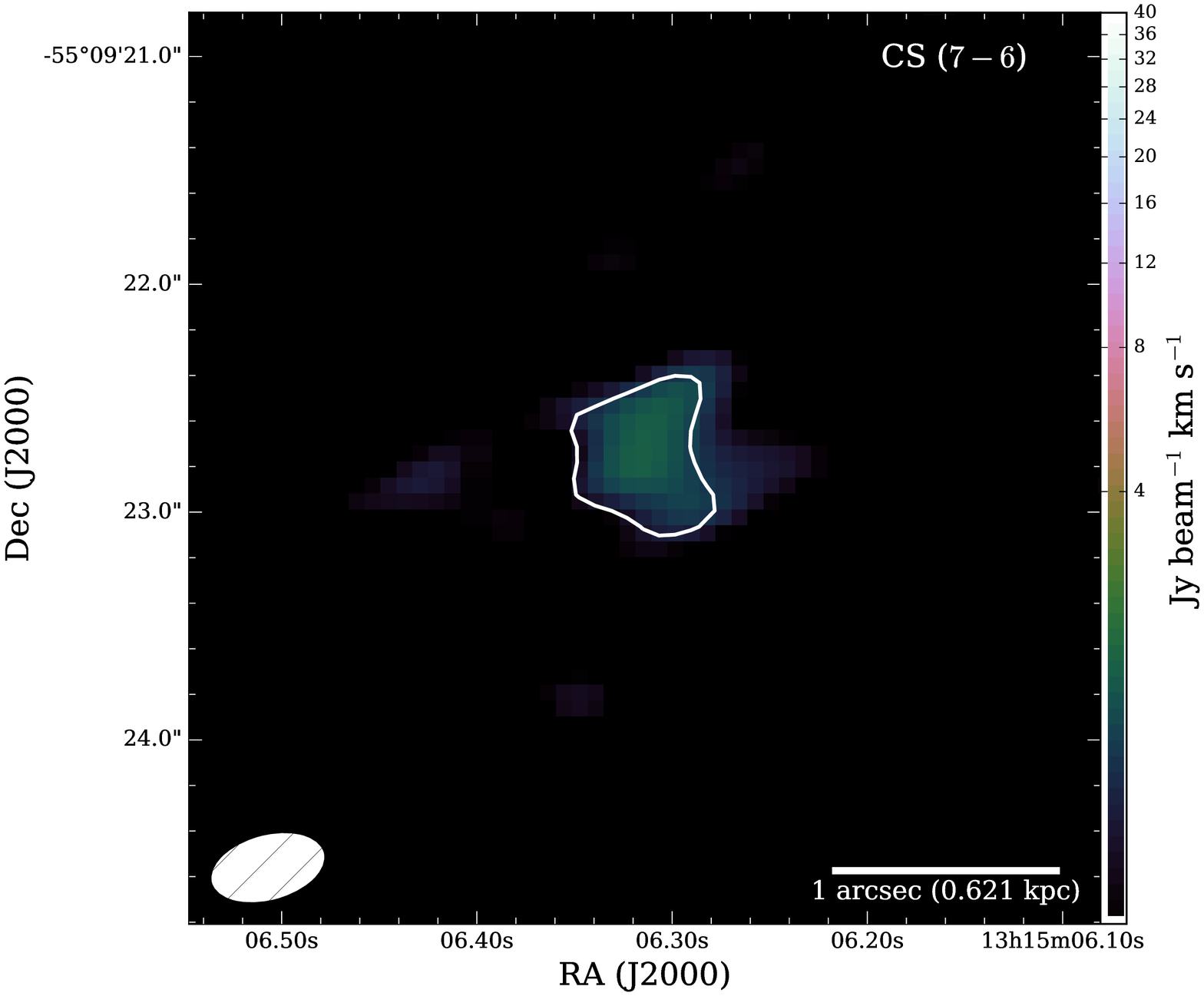}
\includegraphics[width=0.45\textwidth]{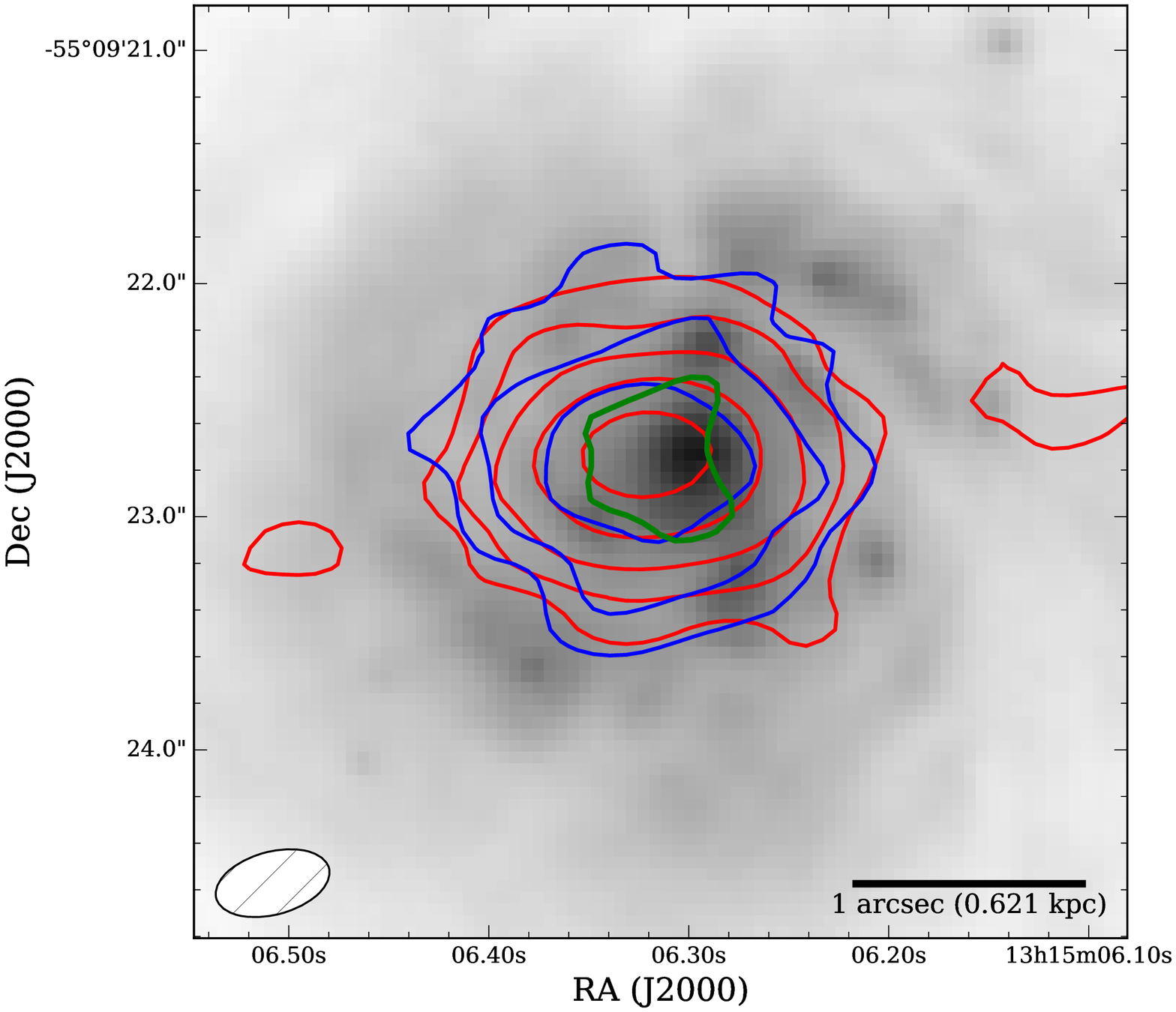}
\caption{Upper left: \HCN{4}{3} total intensity map.
Upper right: \HCO{4}{3} total intensity map.
Lower left: \CS{7}{6} total intensity map.
Lower right: HST/ACS F814W image of the central $\sim7$ kpc of \source, with the total intensity of \HCN{4}{3}, \HCO{4}{3}, and \CS{7}{6} shown in red, blue, and green contours, respectively, to illustrate their relationship to each other and the underlying optical continuum emission.
The three colorscale figures all utilize the same brightness scaling, to illustrate the relative intensity of the three molecular lines.
In all figures, the size of the ALMA synthesized beam is shown in the lower-left.
The emission from these tracers is confined to a molecular disk with an overall extent of $\sim1.2$ kpc.
The contour levels in all 4 panels begin at 1 \Jybeam \kms and increase by factors of 2.
The relative astrometry of the HST/ACS image is uncertain to roughly $1\arcsec$, so the peak of the molecular emission may be consistent with the position of the optical nucleus.}
\label{fig:maps}
\end{figure*}

\begin{figure}
\includegraphics[width=0.47\textwidth]{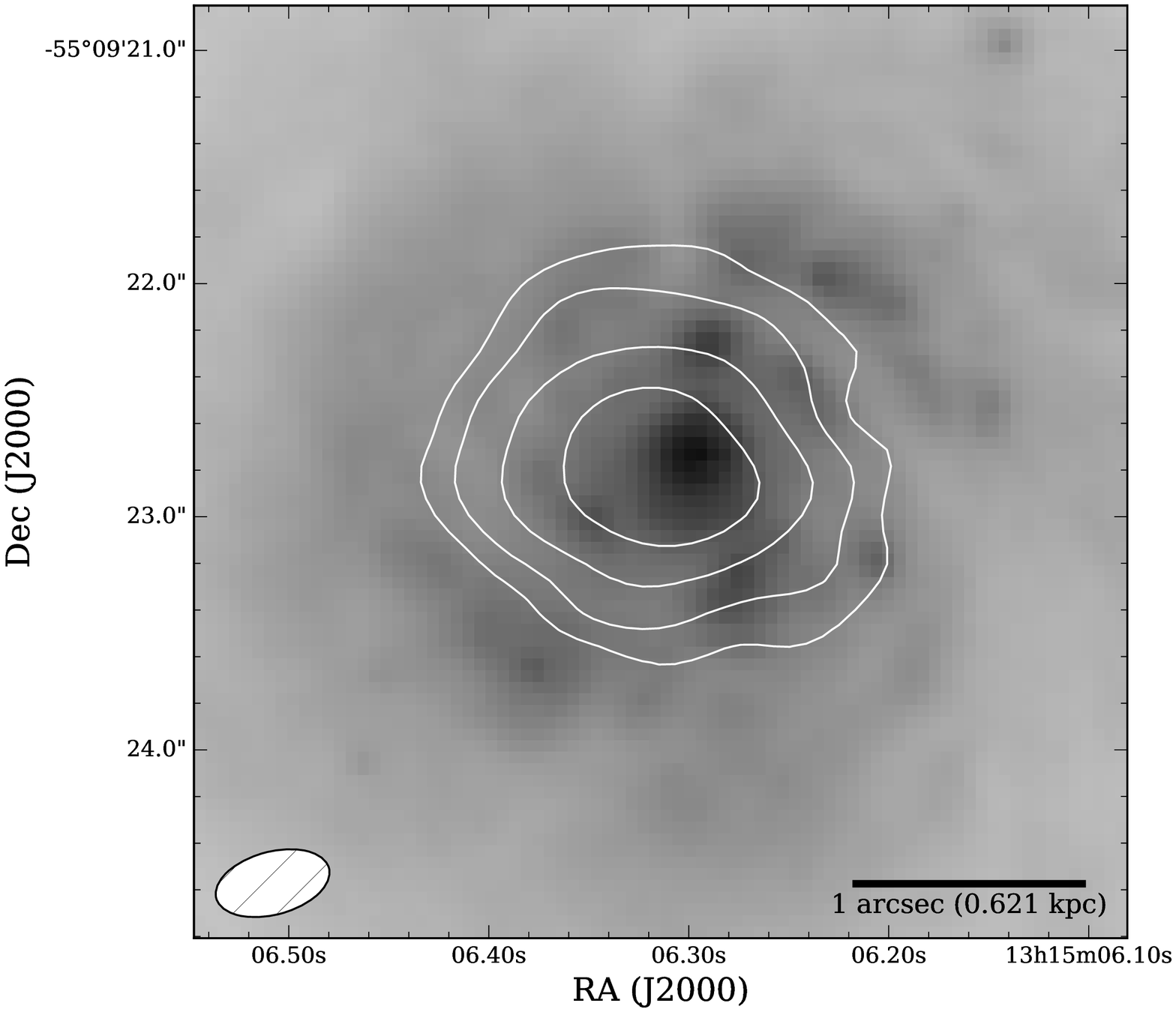}
\caption{Map of the 333 GHz continuum emission (white contours), superimposed on an HST/ACS F814W image of \source.
The contours levels are 1, 2, 4, and 8 \mJybeam.
The relative astrometry of the HST/ACS image is uncertain to roughly $1\arcsec$, so the 0.8 mm continuum peak is consistent with the position of the optical nucleus.
}
\label{fig:continuum}
\end{figure}

Emission from \HCN{4}{3}, \HCO{4}{3}, and \CS{7}{6} were detected at 215, 243, and 20$\sigma$, respectively (Figure~\ref{fig:profiles}).
The emission from these species is compact ($1\arcsec.8$, $1.2$ kpc) and centrally concentrated (Figure~\ref{fig:maps}).
We did not see clear evidence for strong emission of the $v_2=1f$ \HCN{4}{3} vibrational transition, but see Section~\ref{sec:pumping} for discussion.
We also detect the 333 GHz continuum emission at $225\sigma$ (Figure~\ref{fig:continuum}).
Measured parameters are provided in Table~\ref{table:meas}.
Properties of \source derived from these observations are given in Table~\ref{table:derived} and discussed in detail in later sections.

\begin{deluxetable}{lcc}
\tablecaption{Measured Molecular Line and Continuum Properties}
\tablehead{ & \colhead{Integrated Flux} & \colhead{FWHM\tablenotemark{a}} \\ 
& & (\kms) }
\startdata
\HCN{4}{3} & $86.2 \pm 0.4$ Jy \kms & 380 \\
\HCN{4}{3} $v_2=1f$ & $<0.27$\tablenotemark{b} Jy \kms & \nodata \\
\HCO{4}{3} & $48.6 \pm 0.2$ Jy \kms & 360 \\
\CS{7}{6} & \phantom{0}$4.2 \pm 0.2$ Jy \kms & 250 \\
    \hline \\
333 GHz & $89.8 \pm 0.4$ mJy & \nodata 
\enddata
\tablenotetext{a}{Width measured directly from the line profiles.}
\tablenotetext{b}{$1\sigma$ upper limit, assuming a boxcar line with a width of 200 \kms, motivated by the width of detected $v_2=1f$ lines in other systems \citep{Aalto2015a}.}
\tablecomments{Col 1 -- Line identification or continuum frequency, Col 2 -- Integrated flux (for lines) or flux density (for continuum), Col 3 -- Measured full-width of the emission line at half of the observed peak value.}
\label{table:meas}
\end{deluxetable}

\subsection{Comparison with Single Dish Measurements}

\source was observed by \citet{Zhang2014} with the APEX 12m telescope.
Using a conversion of $41$ Jy/K\footnote{Obtained from the APEX website: \url{http://www.apex-telescope.org/telescope/efficiency/}.}, their \HCN{4}{3} and \HCO{4}{3} fluxes are $82\pm12$ Jy \kms and $66\pm12$ Jy \kms, respectively.
Our HCN flux agrees with theirs, suggesting we are recovering the total flux with these ALMA data.
Our HCO$^+$ flux is $\sim25$\% lower than theirs, indicating we may be resolving out some extended flux on scales $\gtrsim8\arcsec$ (the largest recoverable scale for this ALMA configuration and observing frequency) and $\lesssim18\arcsec$ (the beam size of APEX at these frequencies), though emission on scales $\gtrsim4\arcsec$ will also be affected by filtering.
The possible effects of this are further discussed in Section~\ref{sec:missing}.

\citet{Zhang2014} quote an upper limit of $<36.6$ Jy \kms for the \CS{7}{6} line; our measured line flux is nearly a factor of 10 below their upper limit, and thus consistent.

\begin{deluxetable*}{lll}
\tablecaption{Derived Nuclear Properties}
\tablehead{\colhead{Quantity} & \colhead{Value} & \colhead{Units}}
\startdata
$M_{dyn}$ [\HCN{4}{3}, $R<0.5$ kpc]  & $1.2(\sin i)^{-2}\times10^{10}$ & \Msun \\
$M_{dyn}$ [\HCO{4}{3}, $R<0.5$ kpc]  & $1.0(\sin i)^{-2}\times10^{10}$ & \Msun \\
$\Sigma_{dyn}$ [$R<0.5$ kpc]\tablenotemark{a} & $1.2(\sin i)^{-2}\times10^{10}$ & \Msun kpc$^{-2}$ \\
$M_{ISM}$ [total, 333 GHz]\tablenotemark{b}  & $(3.3\pm0.7)\times10^{10}$ & \Msun \\
$M_{ISM}$ [total, 333 GHz]\tablenotemark{c}  & $3\times10^{10}$  & \Msun \\
$M_{ISM}$ [$R<0.5$ kpc, 333 GHz]\tablenotemark{b}  & $(1.4\pm0.4)\times10^{10}$ & \Msun \\
$M_{ISM}$ [$R<0.5$ kpc, \hho modeling]\tablenotemark{d}  & $7.5\times10^{9}$ & \Msun \\
$\Sigma_{ISM,50}$ [within half-light radius of 333 GHz emission] & $5.7\times10^{10}$ & \Msun kpc$^{-2}$ \\
$\Sigma_{IR,50}$\tablenotemark{e} [within half-light radius of 333 GHz emission] & $4.7\times10^{12}$ & \Lsun kpc$^{-2}$
\enddata
\tablenotetext{a}{Calculated using the mean $M_{dyn}$ from HCN and HCO$^+$.}
\tablenotemark{b}{Calculated using the empirical relation from \citet{Scoville2014}. This calibration assumes the \HI mass is equal to $50\%$ of the molecular mass. The \citet{Scoville2016} relation removes the \HI mass from the calibration, resulting in a 1/3 reduction in the inferred mass.}
\tablenotetext{c}{Estimated by computing a dust mass from the 333 GHz continuum emission with the temperature derived from the \hho modeling and assuming a gas-to-dust ratio of 100.}
\tablenotetext{d}{Calculated from the \HH column inferred from the modeling of \hho lines.}
\tablenotetext{e}{Calculated by taking the 80\% of \LIR estimated by \citet{Diaz-Santos2010} to originate within the nuclear starburst, and assuming the \LIR follows the distribution of the sub-mm continuum emission.}
\label{table:derived}
\end{deluxetable*}

Using our detections of several dense gas tracers we investigate the excitation of HCN and HCO$^+$, as well as the spatial variations of the HCN/HCO$^+$ ratio (Section~\ref{sec:excitation}).
These ALMA data further reveal tentative evidence for outflowing dense molecular gas, through wings on the HCN and HCO$^+$ lines (Section~\ref{sec:outflows}).

\subsection{Excitation of HCN and HCO$^+$}
\label{sec:excitation}

\begin{figure*}
\includegraphics[width=0.47\textwidth]{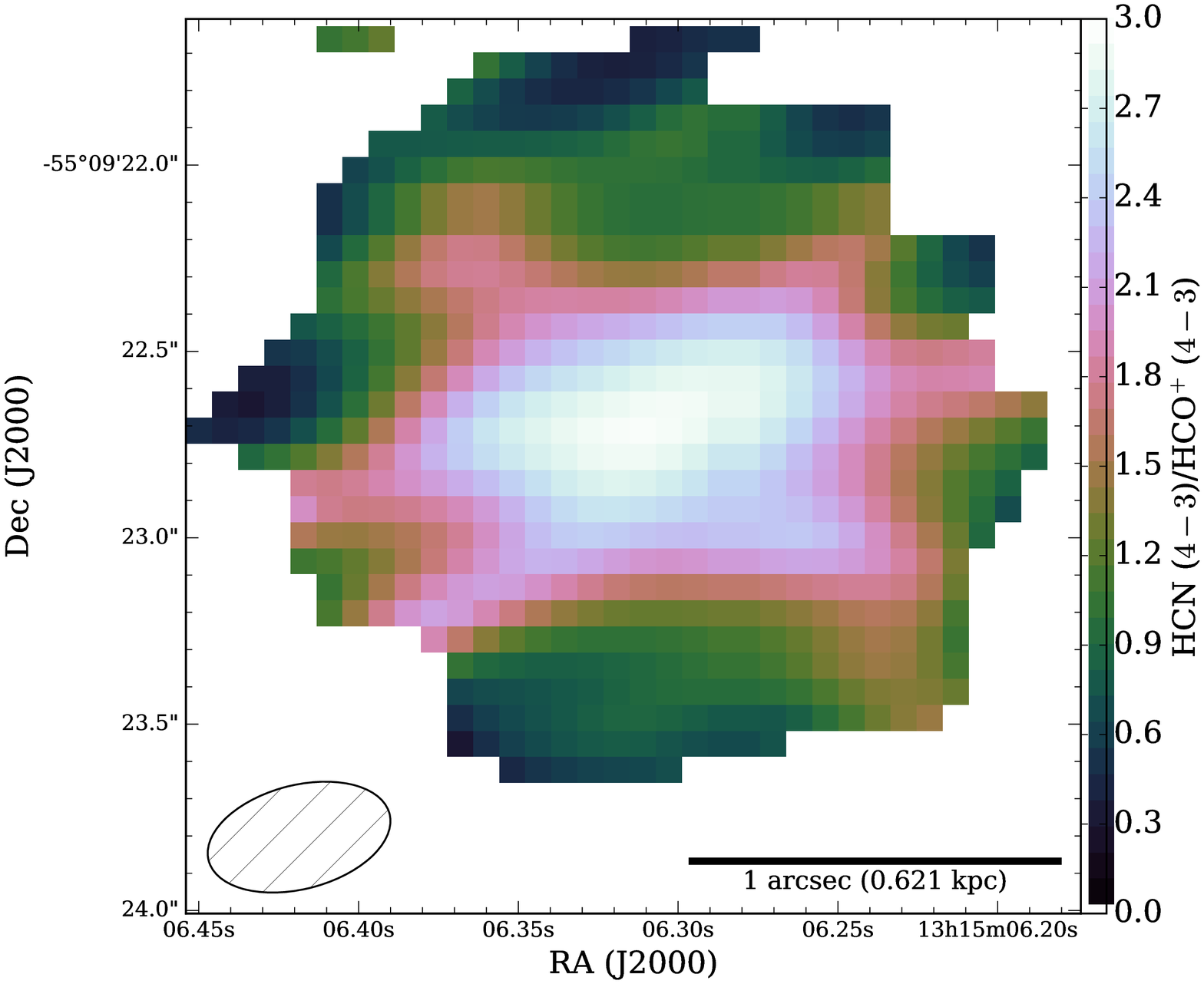}
\includegraphics[width=0.47\textwidth]{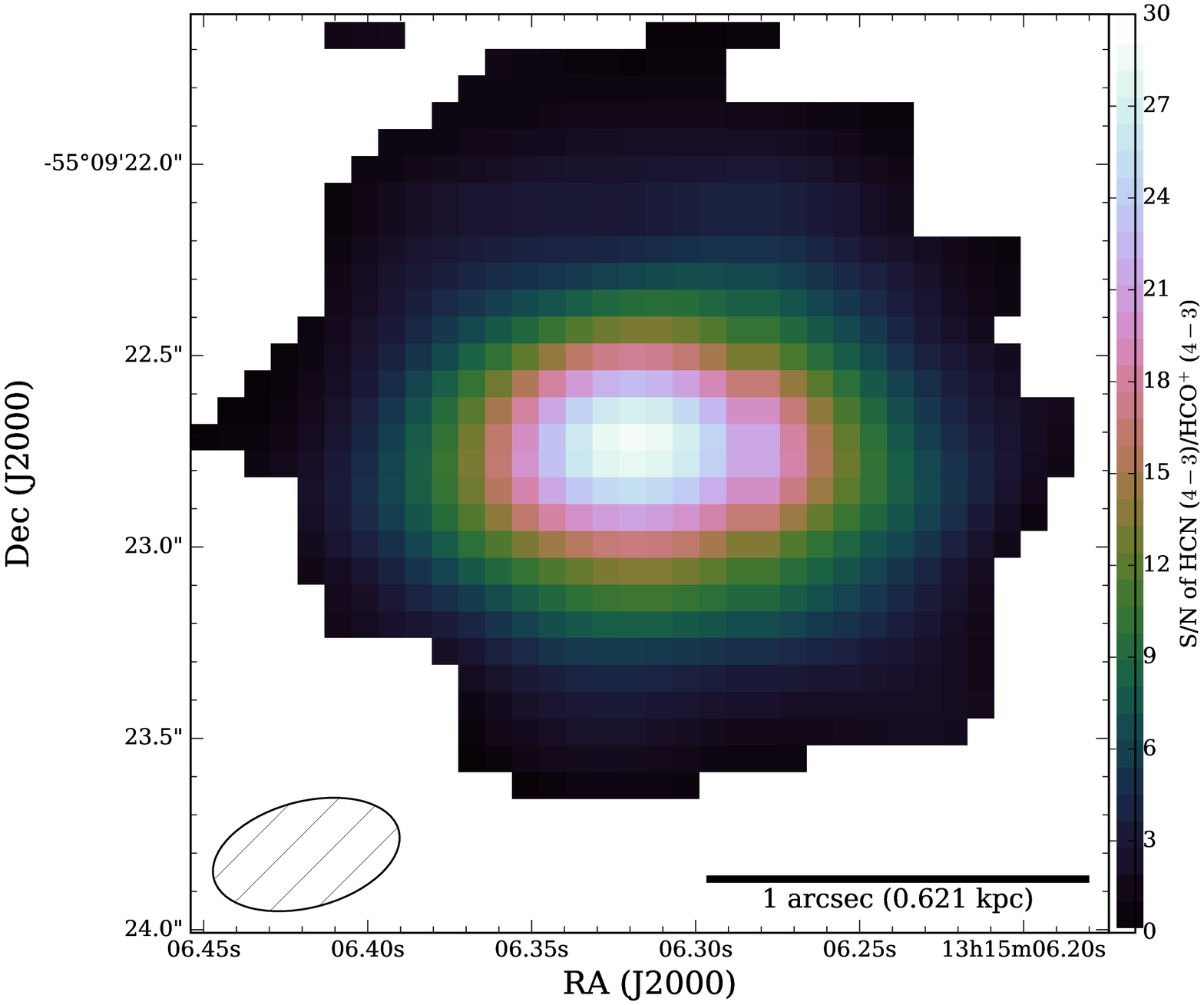}
\caption{Left: Map of the \HCN{4}{3}/\HCO{4}{3} ratio.
Right: Signal-to-noise map for the \HCN{4}{3}/\HCO{4}{3} ratio.
The HCN/HCO$^+$ is high over the nucleus (central $\sim300$ pc) and shows a strong (factor of $\sim3$) decrease in the outer portion of the disk.}
\label{fig:ratio}
\end{figure*}

In Figure~\ref{fig:ratio} we show the spatially-resolved \Lprime{HCN}{4}{3}/\Lprime{HCO$^+$}{4}{3} ratio and its S/N.
The ratio map was created by dividing the total intensity map of \HCN{4}{3} by the total intensity map of \HCO{4}{3}, masking out the regions where \HCO{4}{3} was not detected at $\geq3\sigma$.
We find the HCN/HCO$^+$ ratio peaks at $\sim2.8$ over the nucleus (in the central resolution element: $325\times180$ pc), and decreases to $\sim1$ off the nucleus.
We measure a spatial- and velocity-integrated \HCN{4}{3}/\HCO{4}{3} ratio of $1.77\pm0.01$.

How do these ratios compare with the ratios expected for starburst galaxies?
If we cross-correlate the single-dish \HCN{4}{3} and \HCO{4}{3} measurements of \citet{Zhang2014} with the $6.2~\mu m$ PAH EQW measurements of \citet{Stierwalt2013} and take the PAH EQW as a proxy for mid-infrared AGN dominance (EQW $<0.2$ are dominated by AGN), the star forming galaxies have HCN/HCO$^+$ ratios of between 0.2 and 1.5.
Thus, the extended emission in \source has a ratio consistent with these starburst-dominated systems.
The high HCN/HCO$^+$ ratio over the nucleus may point to different excitation conditions and/or HCN/HCO$^+$ abundance ratios, co-spatial with the AGN and nuclear starburst.
The line ratios for the nucleus are consistent with what is seen for other AGN hosts \citep{Izumi2016}.

We now discuss the potential mechanisms which could plausibly result in an elevated HCN/HCO$^+$ ratio.

\subsubsection{Limits on Vibrational HCN Emission}
\label{sec:pumping}

Rotational-vibrational lines of HCN ($v_2=1f$, J = $4\rightarrow3$ or $3\rightarrow2$) have now been detected in eight galaxies \citep[Aalto et al.\ \emph{in prep}]{Sakamoto2010,Imanishi2013,Aalto2015,Aalto2015a}.
The systems have compact nuclei and high implied infrared luminosity surface densities.
Based on the high infrared luminosity and low \cii/\LFIR ratio \citep[which has been shown to be correlated with starburst luminosity density][]{Diaz-Santos2013}, \source was viewed as a likely candidate for the vibrational HCN lines.

However, we do not detect the $v_2=1f$ \HCN{4}{3} line in \source, with a $3\sigma$ upper limit of $0.81$ Jy \kms, assuming a linewidth of 200 \kms.
We find the $v_2=0 / v_2=1f$ ratio to be $>100$, in contrast to measured ratios of $4-10$ when the $v_2=1f$ line is detected \citep{Aalto2015a}.

The \HCO{4}{3} line has a small ``shoulder'' on the red side (Figure~\ref{fig:outflow}); we interpret this as outflowing dense molecular gas (see Section~\ref{sec:outflows}), but it could plausibly be attributed to the $v_2=1f$ line.
If this feature is in fact the \HCN{4}{3} $v_2=1f$ line, we find a flux of $2$ Jy \kms, which is a factor of $\sim40$ fainter than the main \HCN{4}{3} line.
However, as is evident from the PV diagram for \HCO{4}{3} (Figure~\ref{fig:outflow}), the high-velocity emission is not cospatial with the center of the system, in tension with expectations for emission from a vib-rotational line, which should be centered on the nucleus \cite[e.g.,][]{Aalto2015a}.
Thus, we conclude the line wing on \HCO{4}{3} is \emph{not} the $v_2=1f$ \HCN{4}{3} line.

Is the $14~\mu m$ luminosity surface density high enough to expect appreciable infrared pumping?
Using Spitzer IRS spectroscopy, \citet{Diaz-Santos2010} found \source to have an unresolved core with a size of $\leq2.68$ kpc (FWHM) at $13.2~\mu m$.
Approximately $80\%$ of the mid-infrared emission arises in this core.
Spitzer observations of \source by \citet{Inami2013} show a $14~\mu m$ flux of $F_{14}=0.5$ Jy.
If we take the unresolved portion of the $14~\mu m$ emission and the area we infer from our ALMA observation of the 330 GHz continuum (FWHM of $0.56$ kpc $\times0.49$ kpc; Section~\ref{sec:continuum}), we estimate a $14~\mu m$ luminosity surface density of $\Sigma_{14} \sim 2.6\times10^{11}$ \Lsun kpc$^{-2}$.
This is approximately two to three dex below the lower limit of the $14~\mu m$ surface brightness derived for sources with detected $v_2=1f$ HCN lines in \citep{Aalto2015a}.
Applying an extinction correction to $\Sigma_{14}$ for \source would reduce the discrepancy, but we have no evidence to suggest the mid-infrared emission is being absorbed behind a significant screen of cooler dust.
The significantly lower $\Sigma_{14}$ in \source suggests the $14~\mu m$ continuum may not be effective at radiatively pumping HCN.

Several detections of the HCN $v_2=1f$ lines occur in systems where the HCN and HCO$^+$ emission is strongly self-absorbed, consistent with a scenario in which the nuclear gas has high density and a high column \citep{Aalto2015a}.
There is evidence for some self-absorption in \source, but it is not nearly as significant as seen in CONs with detected $v_2=1f$ emission.
The integrated line profile (Fig.~\ref{fig:profiles}) does show a dip in the center, which could be the result of some foreground absorption.

We have performed some exploratory large velocity gradient (LVG) modeling of the \HCN{4}{3}, \HCO{4}{3}, and \CS{7}{6} lines using the Radex and DESPOTIC codes \citep{vanderTak2007,Krumholz2014}.
We ran grids of models covering a range of densities ($\log_{10} (n / $\cmc$) = 2-7$), column densities ($\log_{10} ($N$_{\textnormal{H$_2$}} /$ \cmcol$) = 21-25$) and relative HCN/HCO$^+$ abundances ($10^{-3}-10^{3}$). 
While the solutions are under-constrained and so we cannot propose ``best'' values for the system, solutions which matched the observed \HCN{4}{3}/\HCO{4}{3} value of 2.8 over the nucleus required relative HCN/HCO$^+$ abundances $\gtrsim10$.
This is similar to the result of \citet{Izumi2016}, who find HCN/HCO$^+$ abundance ratios of a few to $\gtrsim10$ are needed to explain the observed HCN/HCO$^+$ in AGN hosts, while HCN/HCO$^+$ abundance ratios of $\sim1$ can explain the emission in starburst galaxies.
As we will discuss in Section~\ref{sec:activity}, this abundance enhancement is suggestive of mechanical heating from the nuclear starburst.
Measurements of additional transitions of HCN and HCO$^+$ are needed to perform more detailed LVG modeling to simultaneously constrain the \HH density and relative abundance of each species while also constraining the excitation of these tracer molecules.

\subsubsection{Missing Flux and the HCN/HCO$^+$ Ratio}
\label{sec:missing}

What effect does the missing flux in the \HCO{4}{3} line have on our interpretation of line ratios?
\citet{Zhang2014} find a HCN/HCO$^+$ ratio of $1.2\pm0.3$, somewhat lower than what we find here.
If we assume, as a worst-case scenario, that the 17 Jy \kms difference in \HCO{4}{3} flux between our measurements and those of \citet{Zhang2014} was uniformly resolved-out in a $4\arcsec$ region (the scale on which filtering may start to affect these observations), the contribution of this emission to the central resolution element is $0.15$ Jy \kms.
This contribution would only increase the \HCO{4}{3} flux in the central resolution element by $\sim3$\% and so would not substantially affect the ratio at the center of the emission.
At larger radii, where the HCN and HCO$^+$ emission is fainter, the potential contribution is more significant, but still amounts to $\lesssim15$\%.
Thus, we conclude that the missing \HCO{4}{3} flux does not significantly influence our spatially-resolved determination of the HCN/HCO$^+$ line ratio.

\subsection{Dense Molecular Outflows}
\label{sec:outflows}

\begin{figure*}
\includegraphics[width=0.33\textwidth]{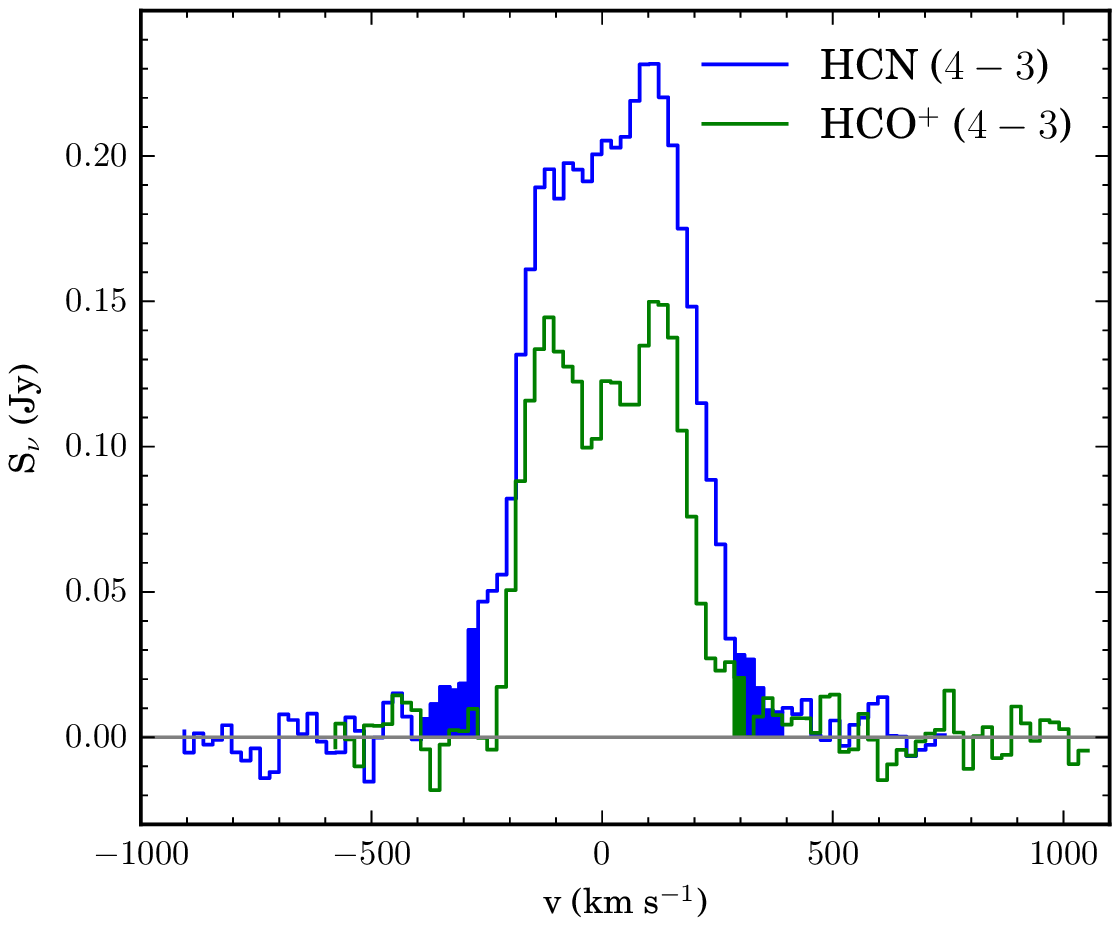}
\includegraphics[width=0.33\textwidth]{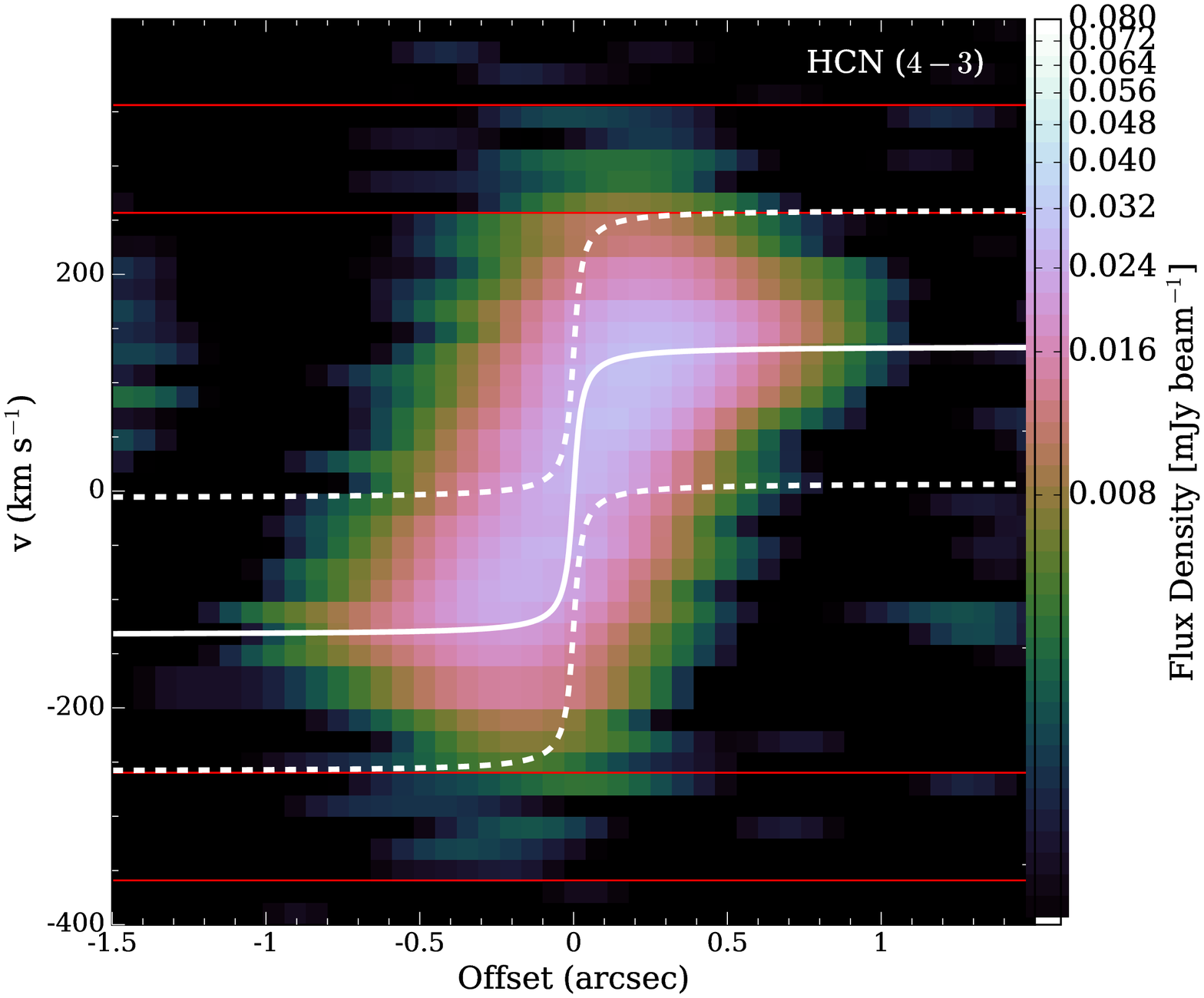}
\includegraphics[width=0.33\textwidth]{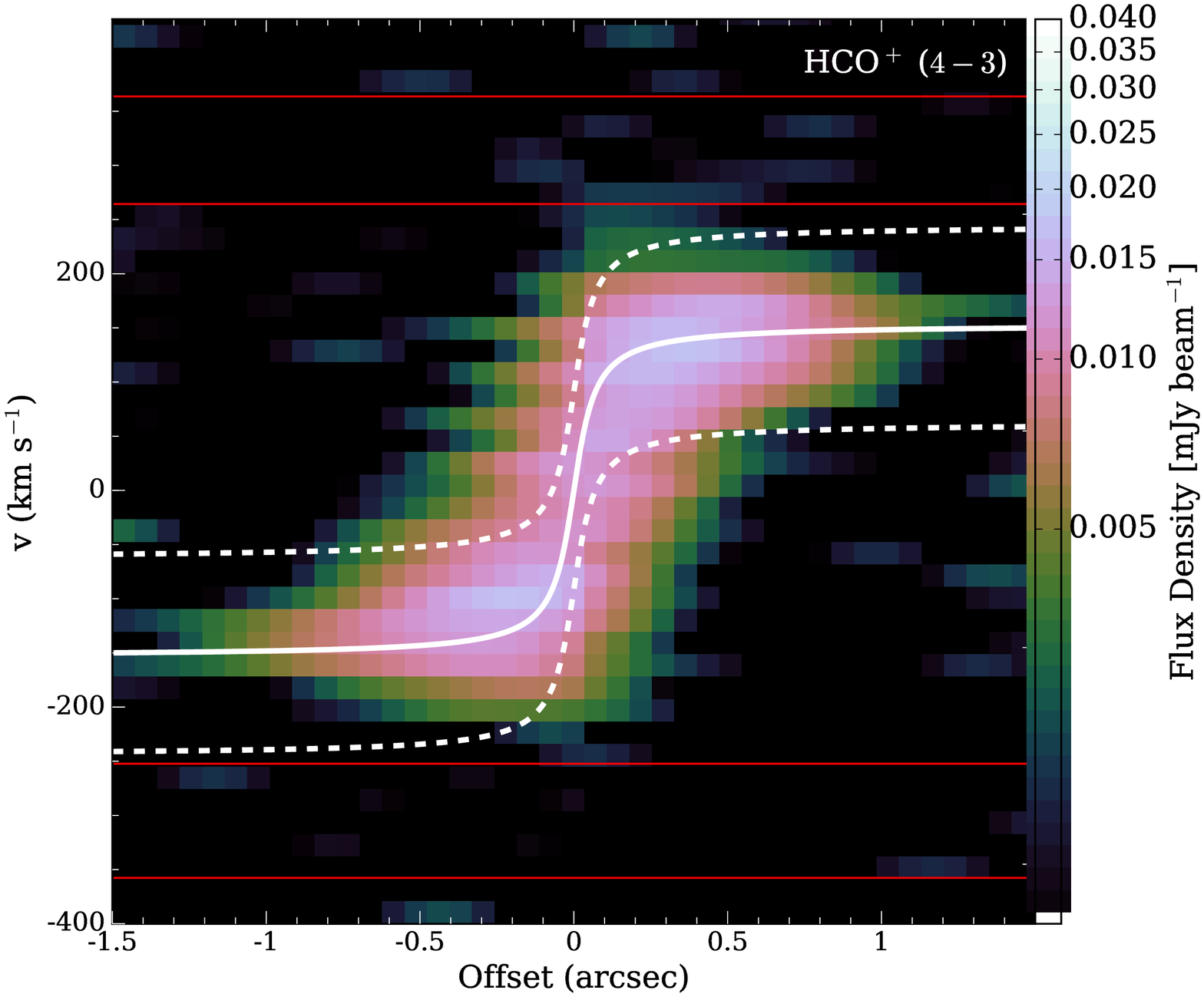}
\caption{Left: Spatially integrated line profiles for \HCN{4}{3} (blue) and \HCO{4}{3} (green).
Here, $0$ \kms corresponds to a redshift $z=0.03112$, based on the location of the \HCN{4}{3} line.
The colored, shaded regions denote the channels associated with outflows.
Approximately 4\% of the HCN flux is in outflowing material, while only 1\% of the HCO$^+$ emission is associated with the outflow.
The HCN emission appears to have both blue and redshifted wings, while the HCO$^+$ emission only has a redshfited wing.
Middle: Position-velocity diagram of \HCN{4}{3}.
Bottom: PV diagram for \HCO{4}{3}.
Both PV diagrams were measured with a cut along the major axis (PA$=94^{\circ}$) of the HCN total intensity map and the data are displayed in a logarithmic scaling.
The emission from both species appears to be mainly confined to the solid-body portion of the rotation curve, though some evidence for flattening is seen at velocities $\sim 200$, particularly in HCO$^+$, where the emission extends over a slightly larger region.
The solid white lines in the middle and right panels show the best-fit rotation curves from GalPak$^{\mathrm{3D}}$ \citep{Bouche2015} modeling.
The dotted lines denote the virial range for the rotation curves.
The red horizontal lines mark the velocity regions where emission is identified as outflows.
This emission identified as winds lies clearly above the flattening of the rotation curve, suggesting it has a non-rotational component to its velocity.
}
\label{fig:outflow}
\end{figure*}

We see wings ($\sim300$ \kms) on both the \HCN{4}{3} and \HCO{4}{3} lines (Figure~\ref{fig:outflow}).
The HCN emission appears to have both blue and redshifted wings while, in contrast, HCO$^+$ appears to only have a small amount of emission in the redshfited wing.
The \HCCCN{39}{38} line lies at an observed frequency of $\sim344.1$ GHz and can possibly contribute to emission on the blue side of \HCN{4}{3}.
However, the imaging of the line wings from the channel shows it is also spatially offset from the nucleus, while we could expect \HCCCN{39}{38} to be centrally concentrated.
Thus we conclude that the line wings are not being contaminated by \HCCCN{39}{38}.

In Figure~\ref{fig:outflow} we show position-velocity diagrams for both species, taken along the observed disk major axis (PA$=94^{\circ}$).
The solid white lines denote the best-fit rotation curve obtained from modeling the data cubes with GalPak$^{\mathrm{3D}}$ \citep{Bouche2015}\footnote{\url{http://galpak.irap.omp.eu/}}.
We also show the expected virial range for the gas \citep[e.g.,][]{GarciaBurillo2015}, defined as a combination of the circular motion, velocity dispersion, and a contribution from in-plane non-circular motion (taken as having a magnitude of $50\%$ of the circular velocity).
The velocities identified as outflows (denoted in Figure~\ref{fig:outflow} by the red lines) lie outside the virial range predicted from the kinematic modeling, indicating they are not participating in quiescent motion within the molecular disk.
The details of the kinematic fits to the \HCN{4}{3} and \HCO{4}{3} lines differ somewhat, this may be due to differential optical depths for the two lines, but they provide consistent estimates for the observed rotation curve and the virial range.
Thus, from Figure~\ref{fig:outflow}, we conclude the emission isolated as outflows is deviating from the rotation curve seen in the gas.

Based on the velocity channels identified as outflows, approximately 4 \% of the detected HCN flux is associated with outflowing gas, while only 1 \% of the HCO$^+$ flux is in outflows.
The luminosity in the outflow is $L^{\prime}_{HCN (4-3),outflow}\approx2\times10^{7}$ K km s$^{-1}$ pc$^2$ and $L^{\prime}_{HCO^+ (4-3),outflow}\approx3\times10^{6}$ K km s$^{-1}$ pc$^2$.
Despite the fact that we are missing 25\% of the single dish HCO$^+$ flux, we are not likely to be missing outflowing gas, given the spatial filtering occurs on scales of of $>4\arcsec$ and due to the fact that the observed HCN outflow is confined to a region only a few synthesized beams across.

The outflow velocities are modest, spanning $200-400$ \kms, in HCN, and $200-300$ \kms in HCO$^+$ (Figure~\ref{fig:outflow}).
With the present data we cannot rule out the presence of dense outflows at the high velocities ($\sim1200$ \kms) seen in OH \citep{Veilleux2013}, though we do not see emission at intermediate velocities.
Other \HCO{4}{3} observations (K.~Sliwa \emph{in preparation}) find blueshifted emission at a velocity of $\sim1100$ \kms, but it is unclear if this is HCO$^+$ associated with the OH outflow or emission from another molecular species.

\subsubsection{Velocity-resolved HCN/HCO$^+$ Ratio}

\begin{figure}
\includegraphics[width=0.47\textwidth]{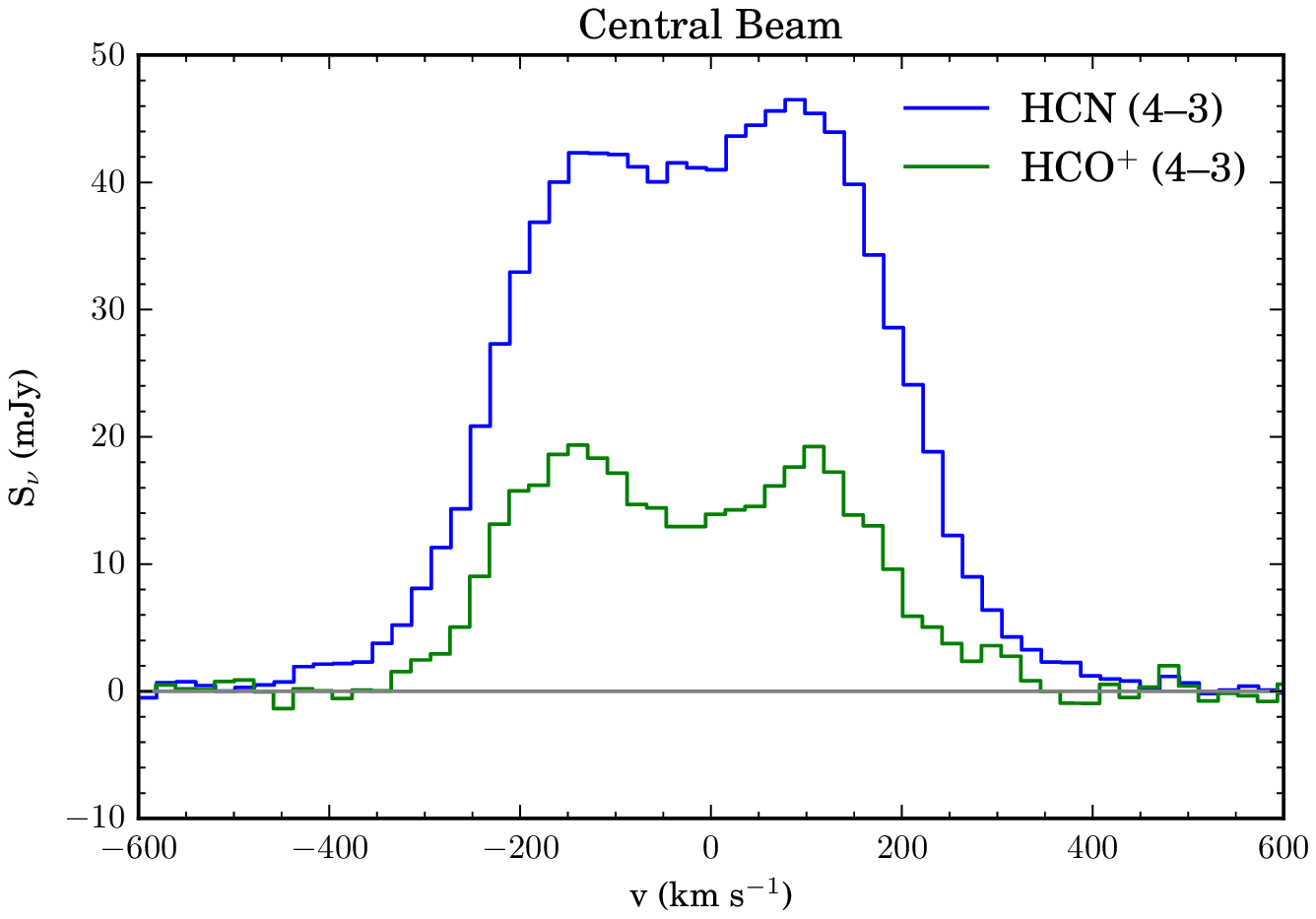}\\
\includegraphics[width=0.47\textwidth]{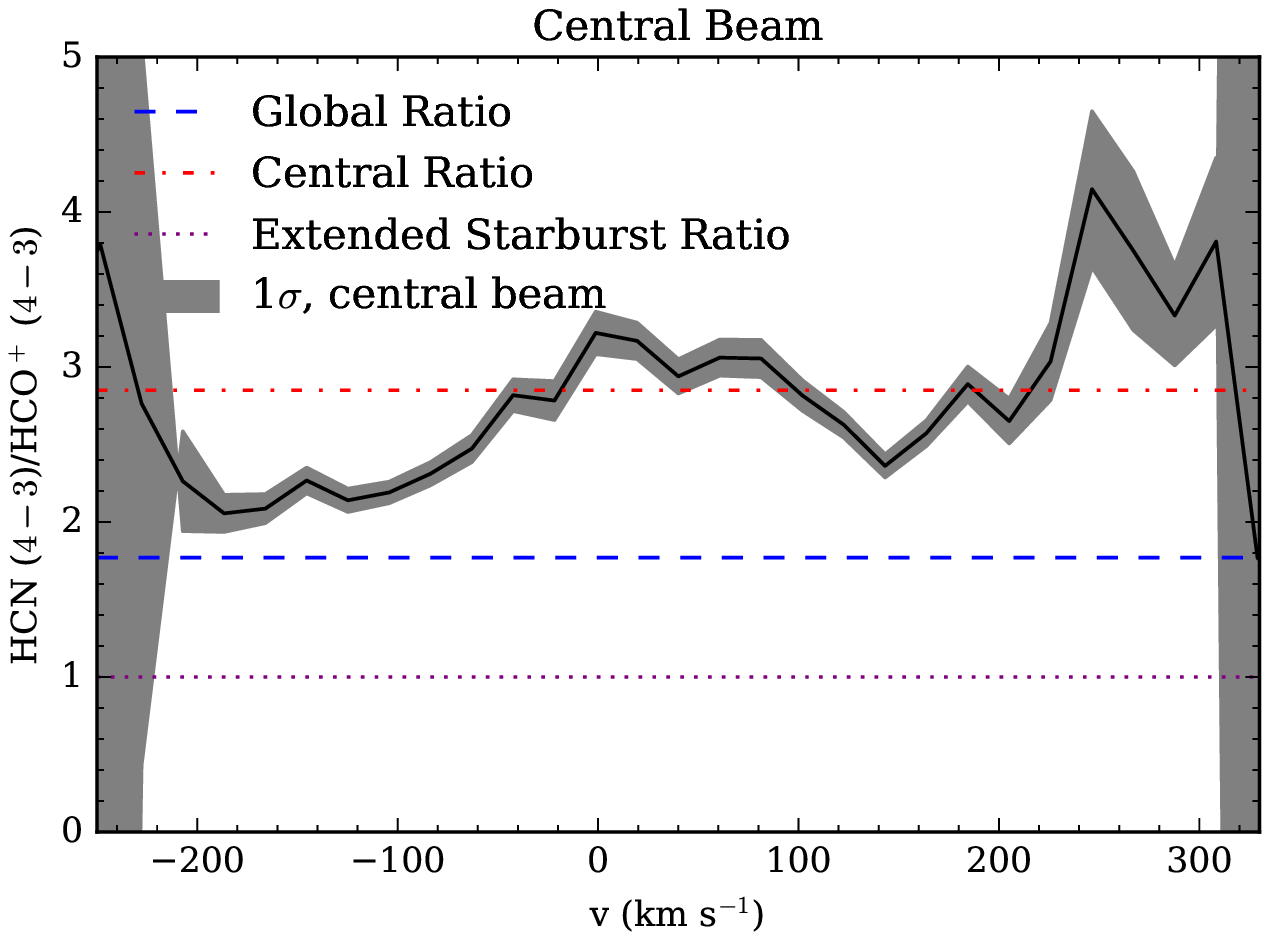}\\
\includegraphics[width=0.47\textwidth]{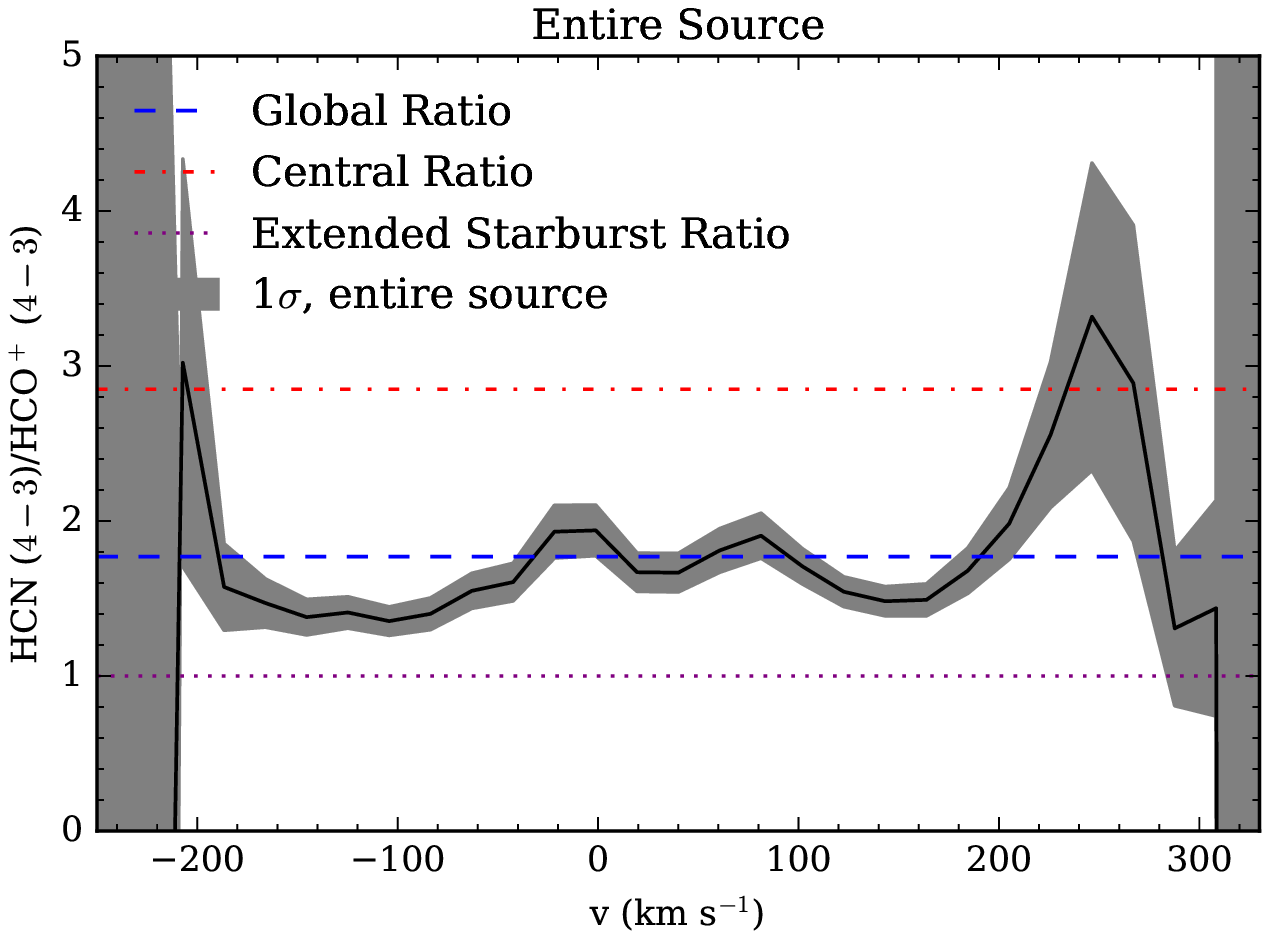}
\caption{
Top: The \HCN{4}{3} and \HCO{4}{3} line profiles in the central resolution element.
Middle: The HCN/HCO$^+$ ratio as a function of velocity, for the central resolution element (black line).
Bottom: The HCN/HCO$^+$ ratio as a function of velocity, integrated over the entire source (black line).
In the lower two panels, the shaded region marks the $1\sigma$ statistical uncertainty in the velocity-resolved ratio.
For comparison we show the velocity-integrated line ratio for the entire source (1.77; blue dashed line) and for the central resolution element (2.85; red dot-dashed line).
The ratio is further enhanced in the channels associated with outflowing gas (Section \ref{sec:outflows}).
The line ratios near the systemic velocity may be affected by self-absorption, which appears to impact \HCO{4}{3} more strongly than \HCN{4}{3}.
}
\label{fig:vel-ratio}
\end{figure}

In addition to positional variations, the HCN/HCO$^+$ ratio varies as a function of velocity both for the entire source (Figure~\ref{fig:vel-ratio}, Middle) and the central resolution element (Figure~\ref{fig:vel-ratio}, Bottom).
In particular, the ratio appears most elevated (HCN/HCO$^+\approx4$) in the high-velocity component of the line, which we attribute to a molecular outflow in the center of the system (Section~\ref{sec:outflows}).
The line ratio in the outflow is approximately the same as the velocity-integrated ratio in the central resolution element, however the line ratio in the central resolution element appears elevated ($>2$) at all velocities (Figure~\ref{fig:vel-ratio} Bottom).
This, in addition to the relatively small contribution of the outflow to the total line flux means the HCN enhancement in the center is not solely due to the presence of the outflow and its high HCN/HCO$^+$ ratio.
The emission from HCN is enhanced over that of HCO$^+$ at all velocities within the central few hundred parsecs of \source.

\section{Water Emission in \source}
\label{sec:water}

\begin{figure}
\includegraphics[width=0.47\textwidth]{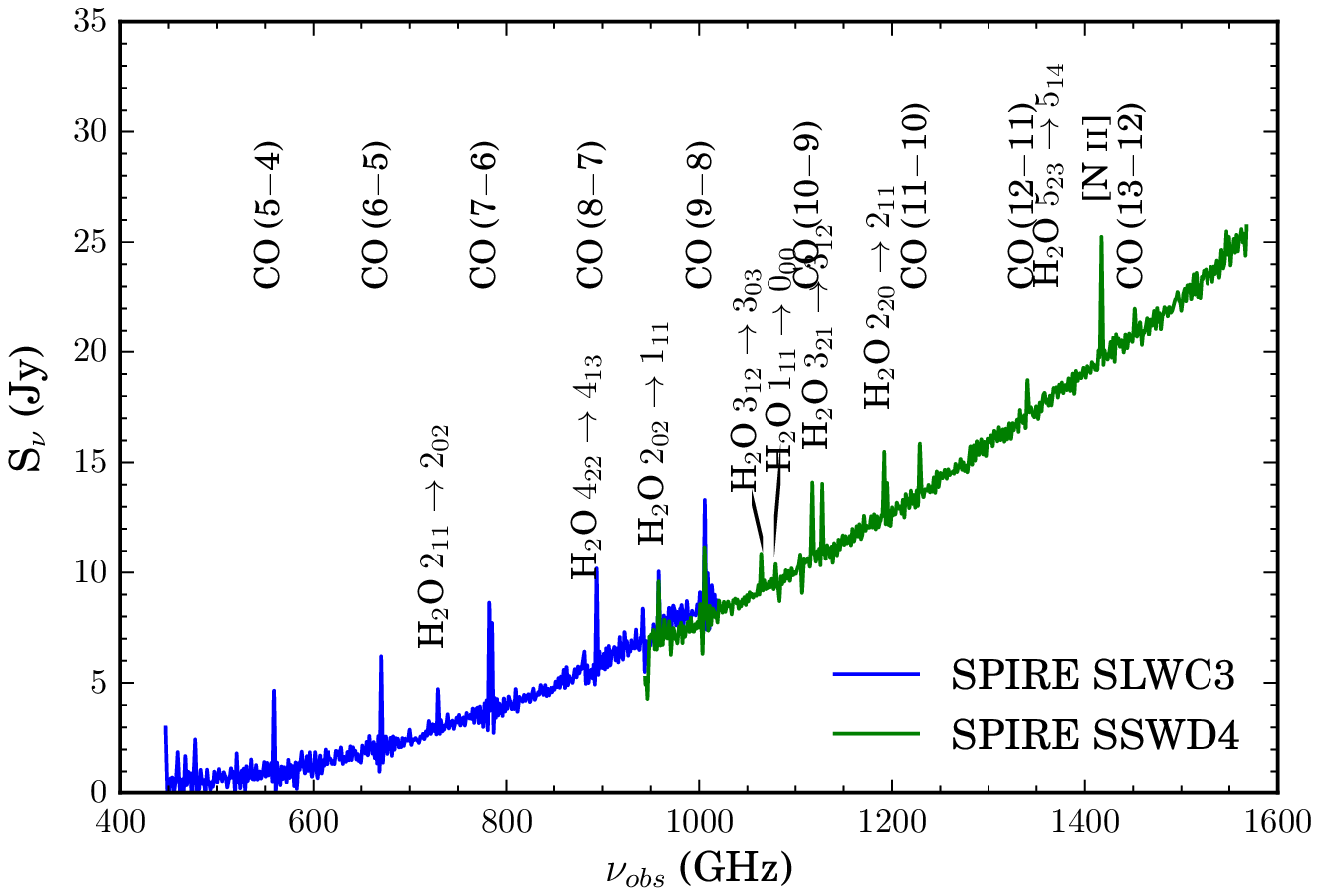}\\
\includegraphics[width=0.47\textwidth]{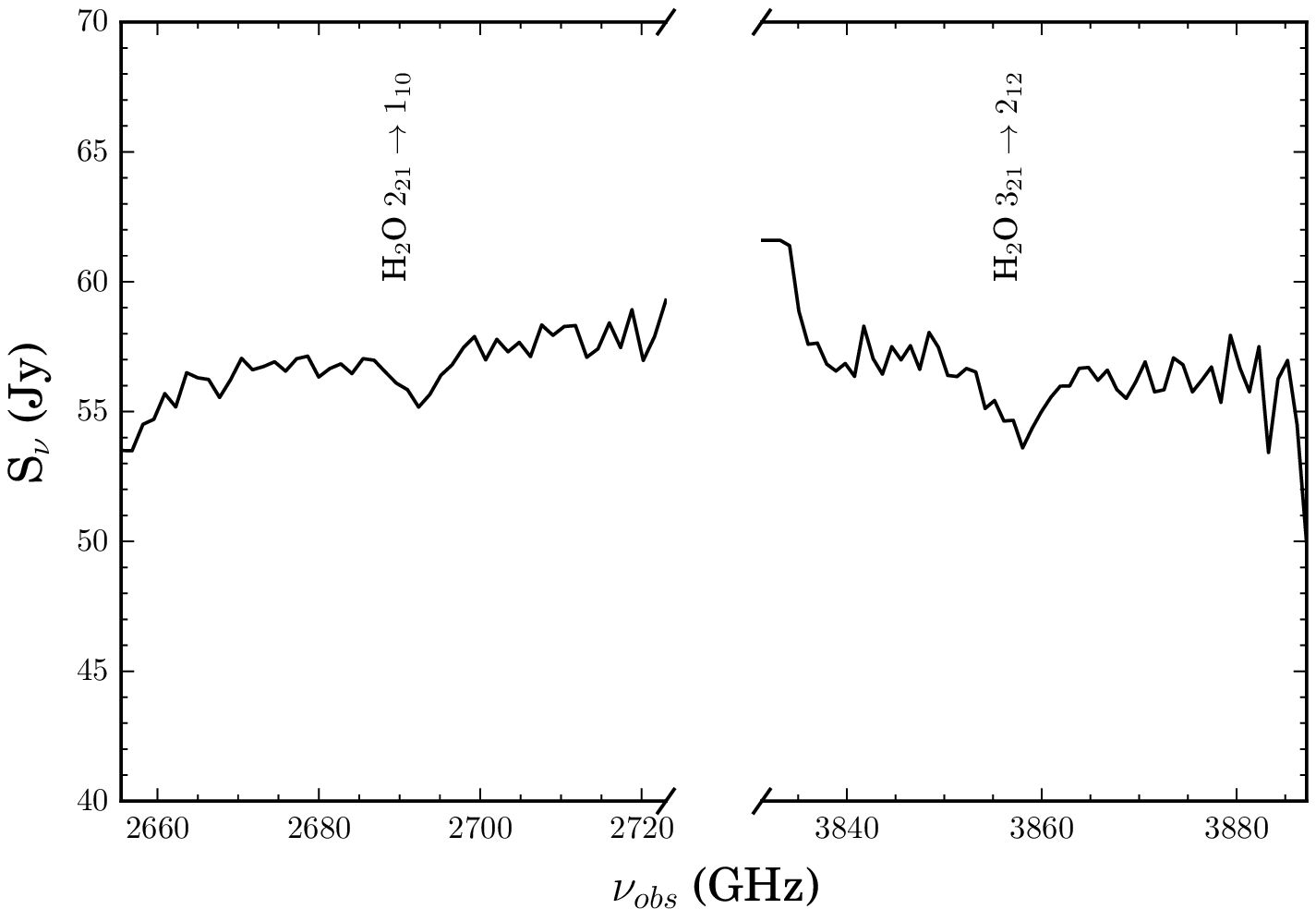}\\
\includegraphics[width=0.47\textwidth]{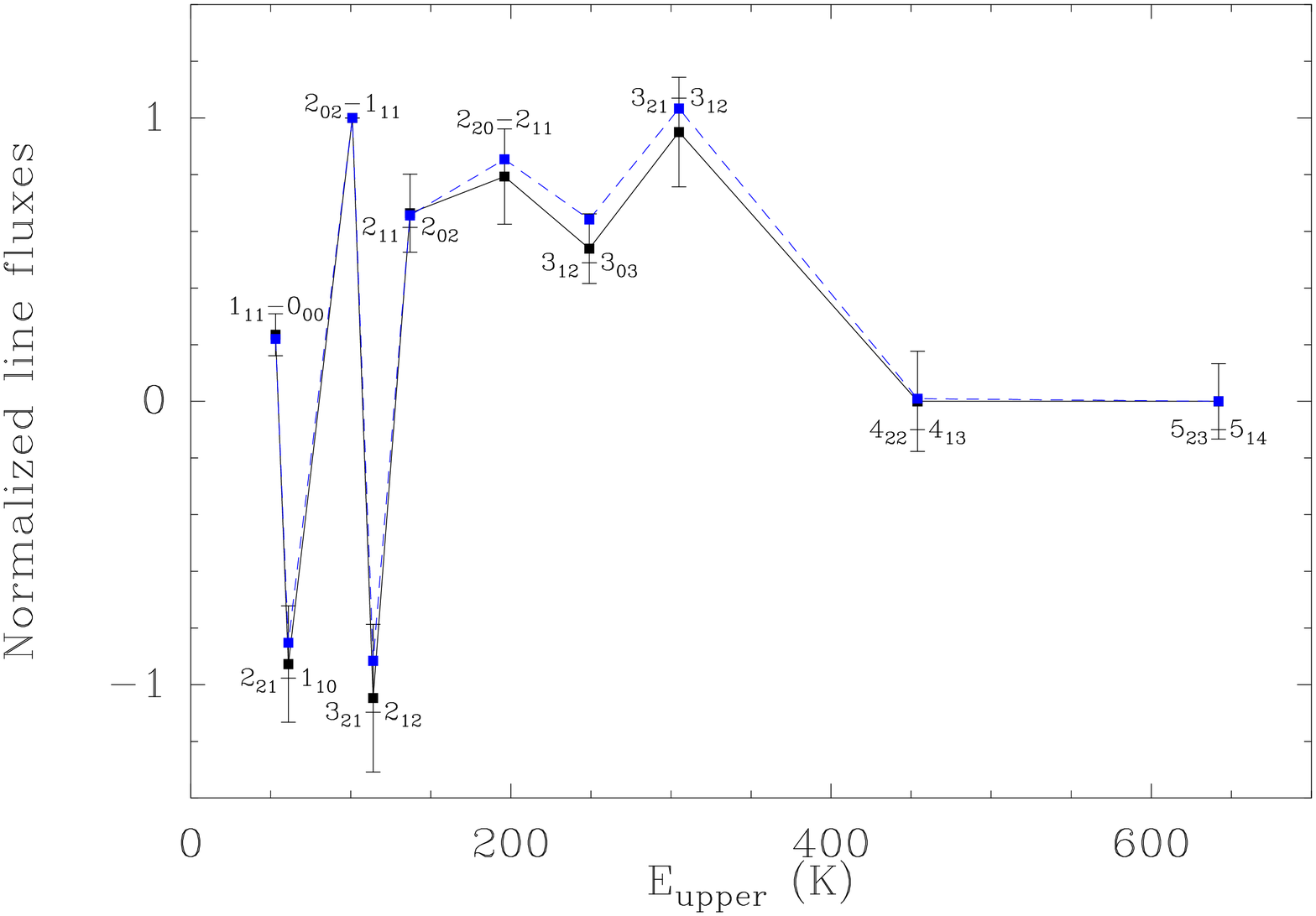}
\caption{Top: Herschel SPIRE spectrum of \source, with the locations of \nii, CO, and \hho lines marked.
Presentation of the CO and \nii lines can be found in \citet{Kamenetzky2015} and N. Lu et~al. (\emph{in preparation}).
Middle: Herschel PACS spectrum, with the locations of \hho lines marked.
See Table~\ref{tab:h2olines} for measured fluxes of the \hho lines and Figure~\ref{fig:herschel} for the \hho SLED.
Bottom: Spectral line energy distribution of the H$_{2}$O lines detected with SPIRE
and PACS.
The black squares present the data, normalized to the flux of the H$_{2}$O $2_{02}\!\rightarrow\!1_{11}$ line. The best fit model is shown as a dashed blue line.
}
\label{fig:herschel}
\end{figure}

Figure~\ref{fig:herschel} shows show the Herschel PACS and SPIRE spectra, as well as the spectral line energy distribution of the \hho lines detected with SPIRE 
and PACS.
The line fluxes are given in Table~\ref{tab:h2olines}.
Two of the ten H$_{2}$O transitions targeted by the Hermolirg project were detected in absorption, both with lower level energies of $\lesssim 100$~K.
We see no obvious contamination by other species in any of the lines.
Five H$_{2}$O transitions, with upper level energies of $\lesssim 300$~K, were detected in emission with SPIRE.

\begin{deluxetable}{lcccc}
\tablecaption{H$_{2}$O lines detected with Herschel}
\tablehead{\colhead{Line} & \colhead{$\nu_{\mathrm{rest}}$} &\colhead{E$_{\mathrm{upper}}$} & \colhead{Cont.\tablenotemark{a}} & \colhead{Flux} \\
& \colhead{(GHz)} & \colhead{(K)} & \colhead{(Jy)} & \colhead{($\mathrm{Jy\,\,km\,\,s^{-1}} $)}}
\startdata
H$_{2}$O $1_{11}\!\rightarrow\!0_{00}$ & 1113.34 & $53$  & $9.6$ & $314.2\pm81.0$  \\
H$_{2}$O $2_{02}\!\rightarrow\!1_{11}$ & \phantom{0}987.93 & $101$ & $6.8$ & $1337.5\pm244.3$ \\
H$_{2}$O $2_{11}\!\rightarrow\!2_{02}$ & \phantom{0}752.03 & $137$ & $2.9$ & $888.1\pm86.2$ \\
H$_{2}$O $2_{20}\!\rightarrow\!2_{11}$ & 1228.79 & $196$ & $12.6$ & $1061.3\pm113.9$  \\
H$_{2}$O $3_{12}\!\rightarrow\!3_{03}$ & 1097.36 & $249$ & $9.2$ & $720.8\pm98.2$ \\
H$_{2}$O $3_{21}\!\rightarrow\!3_{12}$ & 1162.91 & $305$ & $10.8$ & $1271.1\pm113.3$  \\
H$_{2}$O $4_{22}\!\rightarrow\!4_{13}$ & \phantom{0}916.17 & $454$ & $12.0$ & $<236.3$  \\
H$_{2}$O $5_{23}\!\rightarrow\!5_{14}$ & 1410.62 &$642$ & $18.0$ & $<177.8$  \\
\hline
H$_{2}$O $3_{21}\!\rightarrow\!2_{12}$ & 3977.05 &  $114$ & $56.7$ & $-1400.7\pm237$ \\
H$_{2}$O $2_{21}\!\rightarrow\!1_{10}$ & 2773.98 & $61$  & $57.1$ & $-1240.3\pm156$
\enddata
\tablenotetext{a}{Value of the fitted baseline at the line center.}
\label{tab:h2olines}
\end{deluxetable}

We have used the spherically symmetric radiative transfer code described by \citet{Gonzalez-Alfonso1997,Gonzalez-Alfonso1999} to model the observed H$_{2}$O lines and constrain the dust temperature and opacity of the dense ISM.
The code includes collisional excitation as well as excitation by the far-infrared field emitted by warm dust.
Dust is modeled as a mixture of silicates and amorphous carbon, with an adopted mass absorption coefficient as a function of wavelength which is shown in \citet{Gonzalez-Alfonso2014}. 

The models are characterized by the following parameters: the dust opacity at $100$~$\mu$m ($\tau_{100}$), the dust temperature ($T_{\mathrm{dust}}$), the gas temperature ($T_{\mathrm{gas}}$), the H$_{2}$ density ($n_{\mathrm{H_{2}}}$), and the column density of H$_{2}$O per unit velocity dispersion ($N_{\mathrm{H_{2}O}}/\Delta V$).
Collisional rates with H$_{2}$ are taken from \citet{Dubernet2009} and \citet{Daniel2011} for H$_{2}$O.
We have adopted a gas-to-dust ratio of 100 by mass, guided by the average value in LIRGs reported by \citet{Wilson2008}.

Our general approach to the modeling was to compare the observed ratios of various H$_{2}$O lines to a grid of models with varying $T_{\mathrm{dust}}$, $N_{\mathrm{H_{2}O}}/\Delta V$, $\tau_{100}$, $n_{\mathrm{H_{2}}}$, and $T_{\mathrm{gas}}$.
We found that the \hho line ratios in IRAS 13120-5453 cannot be reproduced with collisional excitation alone.
In fact, the excitation is found to be dominated by absorption of photons emitted by warm dust, with collisions mainly affecting the lowest-lying H$_{2}$O levels.
The fact that the H$_{2}$O $1_{11}\!\rightarrow\!0_{00}$ line is detected in emission suggests that some collisional excitation should be taken into account.
We used $T_{\mathrm{gas}}=150$~K and $n_{H_{2}}=3\times10^{4}$~cm$^{-3}$, which yields a thermal pressure similar to that inferred for the warm molecular gas component in Arp~220 \citep{Rangwala2011}.
These values are sufficient to produce emission in the H$_{2}$O $1_{11}\!\rightarrow\!0_{00}$ but still low enough to leave the higher-lying lines relatively unaffected. 

We found that the best fit to the observations is achieved with a dust opacity of $\tau_{100}=0.1-1$, with $\tau_{100}=0.5$ being the preferred value.
For these values of $\tau_{100}$ the relative fluxes of the high-lying lines can be well fitted with different combinations of dust temperatures between $T_{\mathrm{dust}}=40$ and $60$~K, and H$_{2}$O columns between $N_{\mathrm{H_{2}O}}/\Delta V=2\times10^{14}$ and $5\times10^{15}$~cm$^{-2}$ (\kms)$^{-1}$.
The best fit, which is also included in Figure~\ref{fig:herschel}, is achieved with $\tau_{100}=0.5$ $T_{\mathrm{dust}}=40$~K, and $N_{\mathrm{H_{2}O}}/\Delta V=2.5\times10^{15}$~cm$^{-2}$ (\kms)$^{-1}$.

The \hho lines are unresolved with Herschel, but it is likely the emission arises from a similar region to the HCN and HCO$^+$ emission \citep{Gonzalez-Alfonso2014}.
Thus, we can use the observed velocity dispersion of those lines as proxy for $\Delta V$ of the \hho lines (which are unresolved with Herschel).
Taking the mean dispersion of the HCN and HCO$^+$ lines (Figure~\ref{fig:kinematics}) of 140 \kms we find a column density of $N_{\mathrm{H_{2}O}}=3.5\times10^{17}$ \cmcol.
The \HH column inferred from the best-fit $\tau_{100}$ and our assumed dust-to-gas ratio is $\sim6.7\times10^{23}$ \cmcol.
This implies an abundance of \hho relative to \HH of $5\times10^{-7}$, somewhat lower than what is seen in more compact/obscured systems such as Mrk~231 \citep{Gonzalez-Alfonso2008}.

We can compare the \HH column derived from the \hho modeling with N$_H$ determined from modeling of NuSTAR observations.
The NuSTAR $N_H=3.15^{+2.23}_{-1.29}\times10^{24}$ cm$^{-2}$ probes the column between us and the hard x-ray emitting portion of the \source, while the Herschel-derived \HH column probes the entire line of sight through the nucleus (assuming the \hho emission traces the entire ISM).
If the ISM is symmetrically distributed about the nucleus, approximately half of the $\sim6.7\times10^{23}$ \cmcol \HH is between us and the nucleus and the other half is on the far side of the galaxy.
Considering the absorption cross-section of the hydrogen atoms is not significantly affected by being bound in molecules \citep{Cruddace1974,Morrison1983}, the offsetting factors of two imply the molecular ISM could contribute $N_{H,mol}\sim6.7\times10^{23}$ \cmcol, or $\sim20\%$ of the AGN's obscuring column.
The molecular ISM may be a significant contributor to the Compton-thick screen between us and the AGN in \source.

We note the dust temperature derived from modeling of the \hho lines does not provide constraints on the possibility of infrared pumping of HCN as it reflects the dust temperature of the overall nuclear ISM, rather than any compact (10s of pc) hot nuclear core (though such a core appears unlikely to be present, based on the lack of detectable IR pumping and the low $\Sigma_{14}$; Section~\ref{sec:pumping}).

\section{Nuclear Kinematics}
\label{sec:kinematics}

\begin{figure*}
\includegraphics[width=0.47\textwidth]{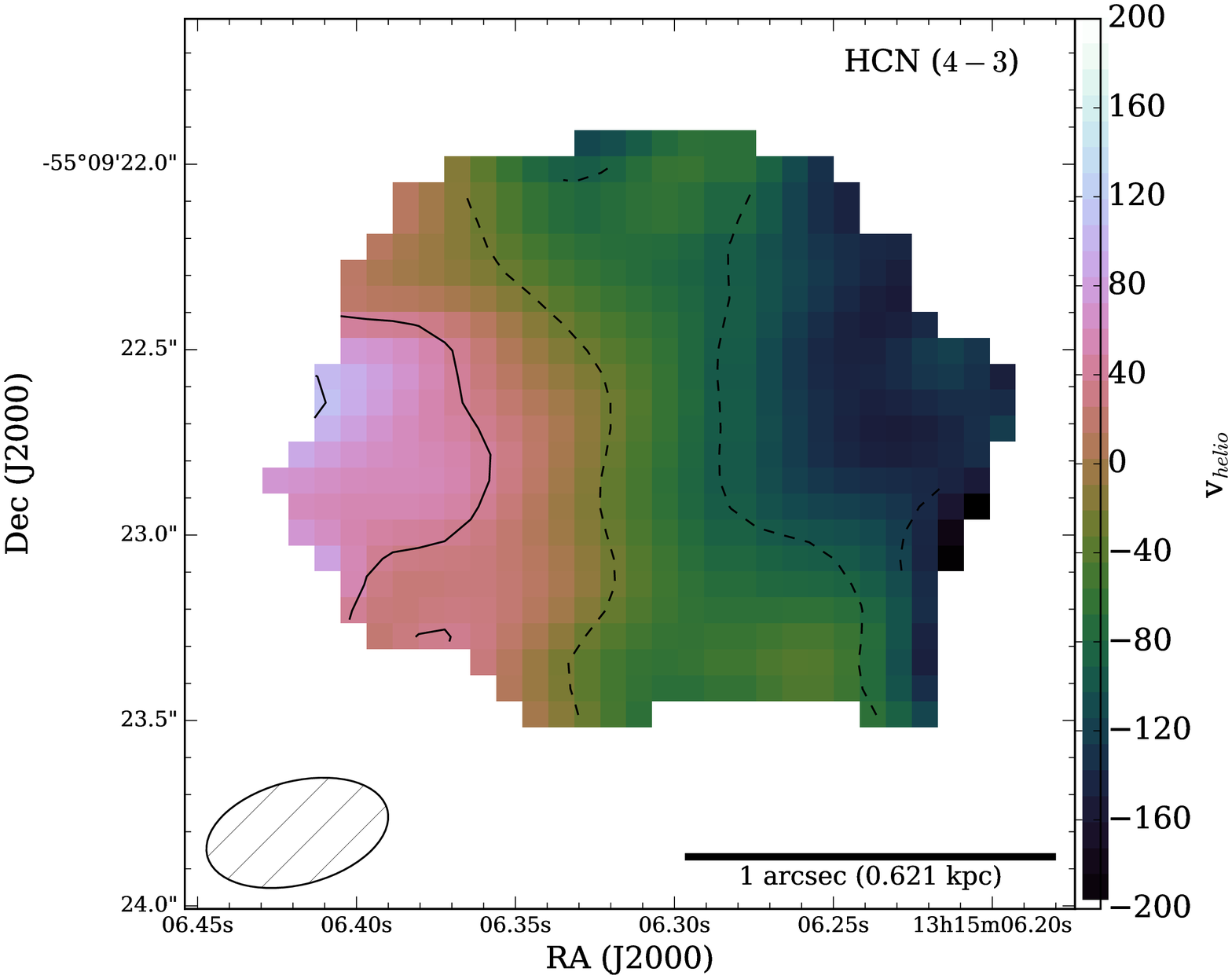}
\includegraphics[width=0.47\textwidth]{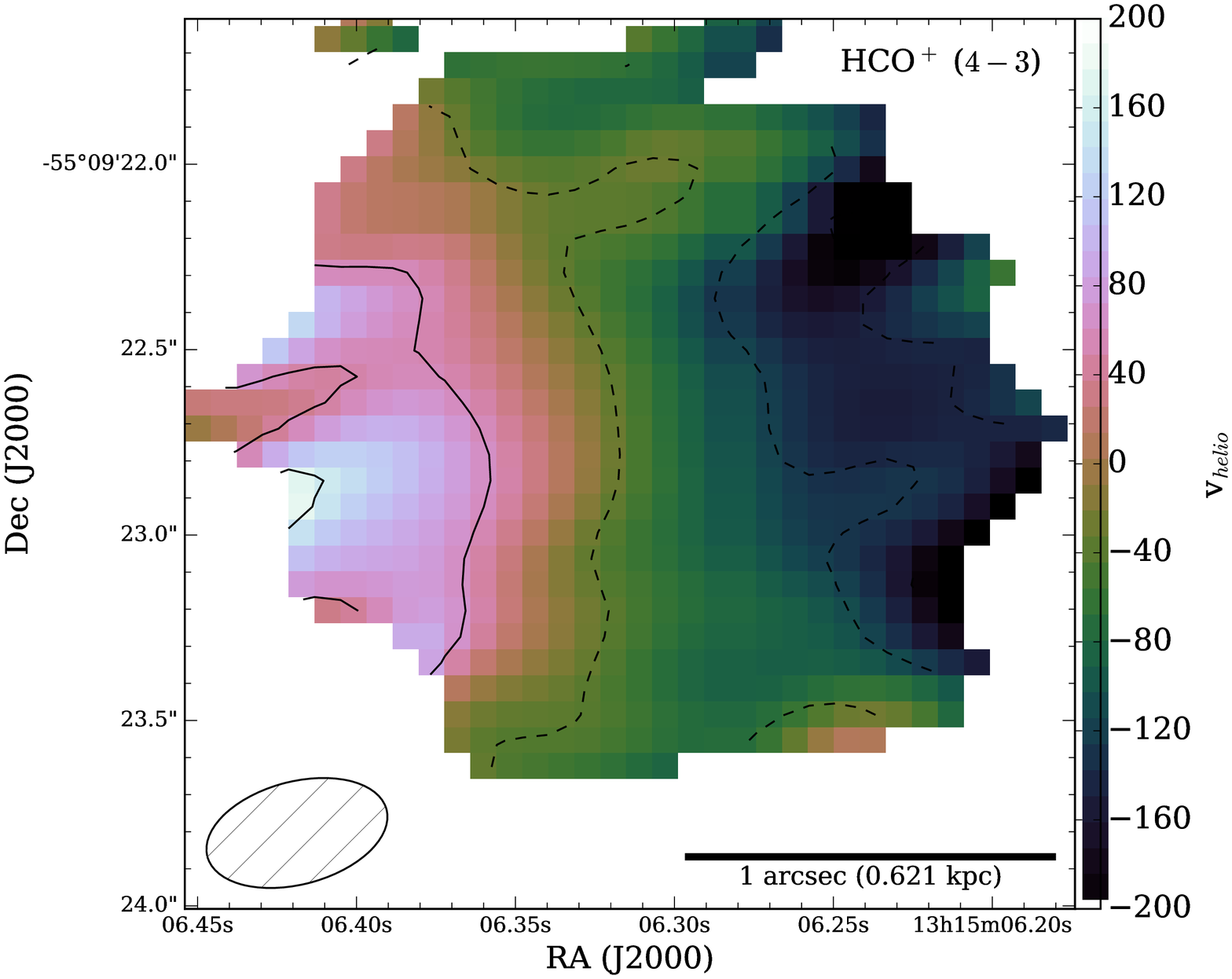}\\
\includegraphics[width=0.47\textwidth]{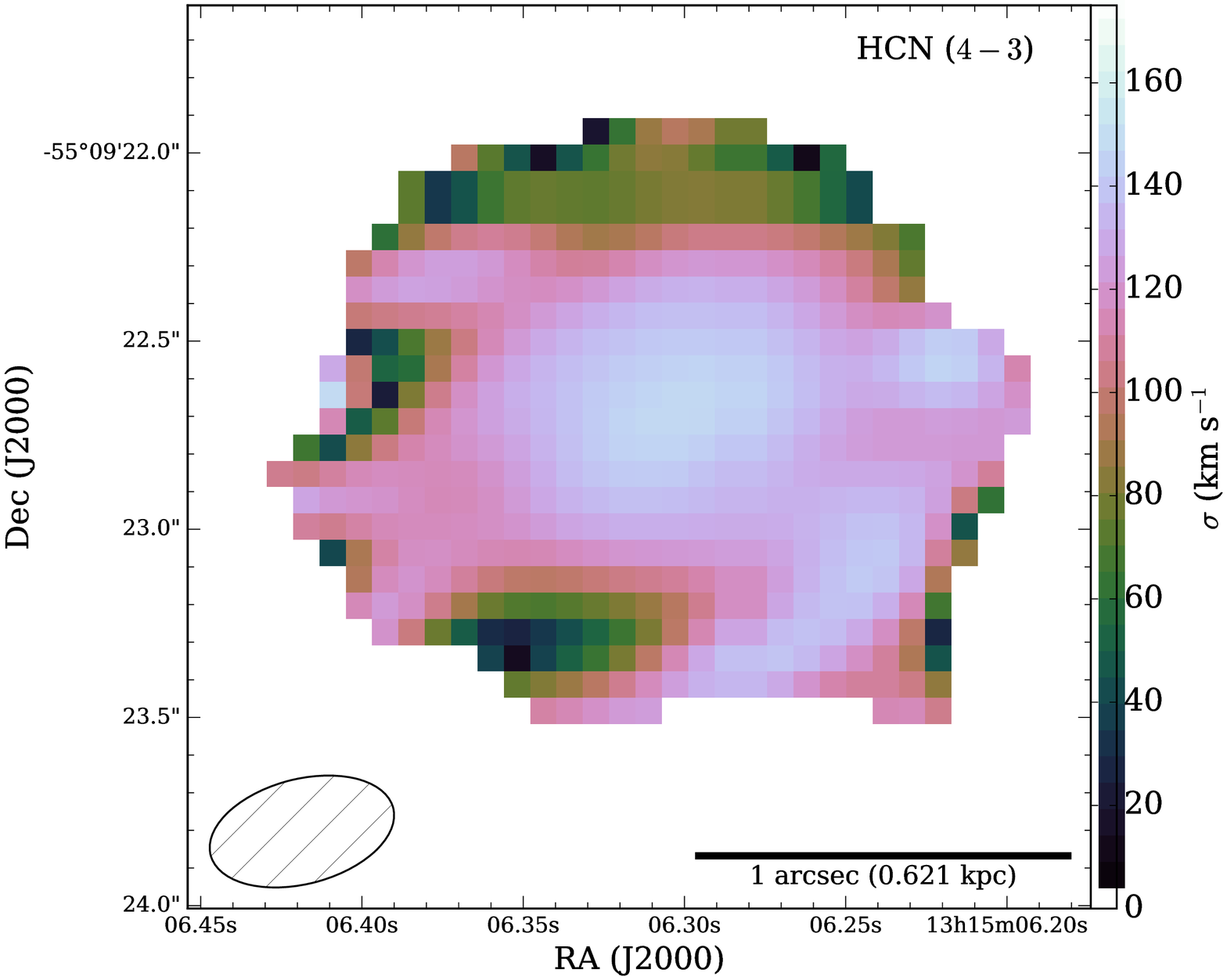}
\includegraphics[width=0.47\textwidth]{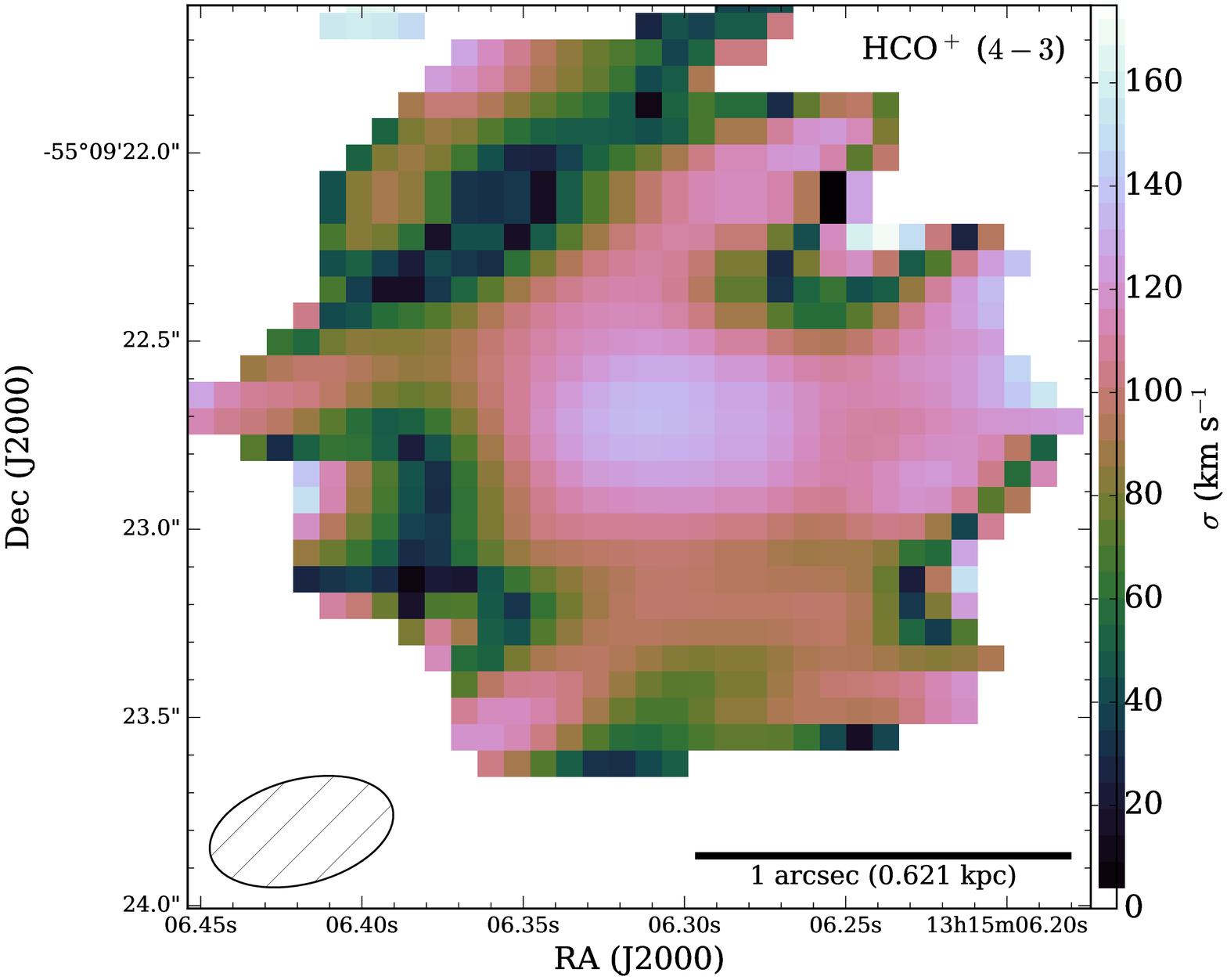}\\
\caption{Top row: Intensity-weighted velocity map for \HCN{4}{3} (left) and \HCO{4}{3} (right).
Contours are spaced every 50 \kms.
Bottom row: Intensity-weighted velocity dispersion map (moment 2) for \HCN{4}{3} (left) and \HCO{4}{3} (right).
The extent of each map was determined by masking the datacube at the $3\sigma$ level in the corresponding total intensity (moment 0) map.
The velocity fields of both molecules are consistent with ordered rotation, in a $\sim1$ kpc central molecular disk.
Both species show centrally peaked velocity dispersions, though the HCO$^+$ peaks at a slightly lower value.
}
\label{fig:kinematics}
\end{figure*}

In Figure~\ref{fig:kinematics} we show the intensity-weighted velocity field (moment 1; top row) and velocity dispersion (moment 2; bottom row) for \HCN{4}{3} (left) and \HCO{4}{3} (right).
The emission from both species appears broadly consistent with a rotating disk, with centrally-peaked velocity dispersion.

Correspondingly, the PV diagrams taken along the observed major axis of the emission (Figure~\ref{fig:outflow}) show clear signatures of rotation.
The (4--3) emission from both species appears to be mainly concentrated in the solid-body rotation portion of the potential, though the \HCO{4}{3} velocity begins to flatten out beyond $\sim0\arcsec.75$ ($0.48$ kpc).
The emission tentatively identified as the outflow lies above the apparent turnover in the rotation curve and outside the estimated virial range, consistent with a scenario in which the gas motion is not solely due to the gravitational potential.

We adopt $z=0.03112$ as the systemic redshift, based on the moment 1 value at the center of the detected ALMA continuum emission (13h15m06.32s, --55d09m22.78s).
This is somewhat higher than the optical redshift found on NED ($z=0.030761$) as measured by \citet{Strauss1992}.
The optical redshift was determined using \Ha, and so could be somewhat offset from the true systemic by obscuration.

\subsection{Nuclear Dynamical Mass}
\label{sec:dynamicalmass}

Using both the \HCN{4}{3} and \HCO{4}{3} emission, we can estimate the dynamical mass as $M\sim V_{circ}^{2}R/G$.
Within a 1 kpc diameter region, HCN and HCO$^+$ have FWHMs of 385 \kms and 375 \kms, respectively and we compute the circular velocity as $V_{circ}=FWHM/(2\sqrt{\ln 2})$.
This corresponds to estimated dynamical masses of $1.2(\textrm{sin}~i)^{-2}\times10^{10}$ \Msun and $1.0(\textrm{sin}~i)^{-2}\times10^{10}$ \Msun.
Adopting the mean value of $1.1(\textrm{sin}~i)^{-2}\times10^{10}$ \Msun, and the area ($0.78$ kpc$^{-2}$), we find a total mean mass surface density of $1.4(\textrm{sin}~i)^{-2}\times10^{10}$ \Msun kpc$^{-2}$.
Assuming the disk is intrinsically circular, the observed axis ratios suggest $i\approx55^{\circ}$, leading to an inferred dynamical mass of $1.6\times10^{10}$ \Msun and a mean mass surface density of $2.0\times10^{10}$ \Msun kpc$^{-2}$ within the central kpc.

\section{Nuclear Continuum Emission}
\label{sec:continuum}

We detect $333$ GHz continuum emission with a total flux of $89.8\pm0.4$ mJy.
For this flux, and $T_{dust}=40$ K derived from the \hho modeling, and a mass absorption coefficient of the dust at 333 GHz of $\kappa(333~\textnormal{GHz})=0.28$ cm$^{2}$ g$^{-1}$ \citep[consistent with Milky Way dust properties;][]{Bianchi2013}, this implies a dust mass of $2.9\times10^{8}$ \Msun.
If we assume a gas-to-dust ratio of 100 \citep{Wilson2008}, the ISM mass is then $\sim3\times10^{10}$ \Msun.
Alternately, using the empirical $L_{850~\mu m}-M_{ISM}$ relation derived from low-z galaxies \citep{Scoville2014}, the continuum emission implies a similar total ISM mass of $(3.2\pm0.7)\times10^{10}$ \Msun.\footnote{We note the \citet{Scoville2014} calibration includes the \HI mass and assumes this mass of this atomic ISM component is equal to $50\%$ of the molecular mass.
The \citet{Scoville2016} calibration does not consider the \HI mass, so the empirical normalization is reduced by 1/3.
Thus, to obtain only the molecular ISM mass (equivalent to the \citealt{Scoville2016}) calibration, ISM mass derived directly from the continuum flux should be multiplied by $0.67$.}

The detected continuum emission is contained within an ellipse with axes of $2.5\arcsec \times 1.9\arcsec$ ($1.6$ kpc $\times~1.2$ kpc).
The distribution of the emission is well represented by a 2D Gaussian with a beam-deconvolved size (FWHM) of $0.86\arcsec \times 0.75\arcsec$ (0.56 kpc $\times$ 0.49 kpc) at a PA of $60^{\circ}$, and is marginally resolved.
This is the best measurement for the size of the nuclear starburst in \source, improving the constraints on the area of the starburst by a factor of $\sim25$.

\citet{Diaz-Santos2010} estimated that $\sim80\%$ of \LIR originates in the compact starburst.
If we assume \LIR is distributed in the same way as the sub-mm continuum emission, we can estimate the IR luminosity surface density within the half-light radius as: $\Sigma_{IR,50}=L_{IR,50}/A_{50}$, where $L_{IR,50}$ is the luminosity contained within the half-light radius of the nuclear starburst ($0.4$\LIR, considering only the unresolved portion) and $A_{50}=\pi a_{sb}b_{sb}$ is the half-light area with semi-major and semi-minor axes of $a_{sb}$ and $b_{sb}$.
We find $\Sigma_{IR,50}=4.7\times10^{12}$ \Lsun kpc$^{-2}$.
This is $\sim10$x larger than the $\Sigma_{IR}$ inferred by \citet{Diaz-Santos2010} based on Spitzer data.
Comparing to the \cii deficit relation for starbursts from \citet{Diaz-Santos2014}, our $\Sigma_{IR,50}$ places \source on the correlation of the \cii deficit and starburst luminosity surface density, suggesting the \cii deficit in \source can be mostly explained by the compact starburst and does not require significant AGN contribution to \LFIR.
Using the same size, we find an ISM surface density of $\Sigma_{ISM,50}=5.7\times10^{10}$ \Msun kpc$^{-2}$ within the half-light radius.

Within a 1 kpc diameter, the 333 GHz continuum emission is consistent with an ISM mass of $(1.4\pm0.4)\times10^{10}$ \Msun and a corresponding mass surface density of $\Sigma_{ISM,1~\textnormal{kpc}}=(1.7\pm0.5)\times10^{10}$ \Msun kpc$^-2$.
This is comparable to the dynamical mass estimated from the HCN and HCO$^+$ kinematics (Section~\ref{sec:dynamicalmass}) and implies $M_{ISM}/M_{dyn}\approx0.9$ within the central kpc.
It is possible the ISM mass estimate from these ALMA data, which use the empirical calibration of \citet{Scoville2014}, are biased high, relative to the galaxies used to calibrate it if the mass-weighted dust dust temperature is higher in \source.
The empirical calibration was arrived at using a galactic conversion factor between L$_{CO}^{\prime}$ -- if the conversion factor varies from this value in the galaxies used to calibrate the relation, the resulting empirical relationship would overestimate the ISM mass.

Alternately, the gas-to-dust ratio may be different in \source, compared to the calibration sources in \citet{Scoville2014}.
\citet{Wilson2008} found a mean gas-to-dust ratio of $120$ for IR-luminous galaxies, but there is an order of magnitude range in the gas-to-dust ratio for the systems in that study ($(29\pm8)-(725\pm286)$).
If the gas-to-dust ratio in \source is on the lower end of the range, the actual ISM mass would be correspondingly lower and in less tension with the dynamical mass.
Furthermore, our ISM mass estimate calibration includes an assumed \HI component; if we instead adopt the calibration of \citet{Scoville2016} for only the molecular ISM, we would find a value reduced by $1/3$.
For comparison, the nuclear gas fraction in systems such as Arp~220 and NGC~6240 are $\sim$1/3 \citep{Scoville1997,Solomon1997,Sakamoto1999,Downes2007,Scoville2015}, showing that M$_{ISM}$ is a substantial fraction of M$_{dyn}$ in ULIRGs.
If we assume \source has a similar gas fraction, the implied gas-to-dust ratio would be on the order of 30 or 40, consistent with the range of \citet{Wilson2008}, but on the lower end.
Finally, the ISM mass estimate derived from the dust mass is sensitive to the choice of $\kappa$.
An elevated value of $\kappa$ would reduce the calculated dust mass and the inferred ISM mass.

We can make an additional estimate of the ISM mass from the \HH column density, $N_H$, inferred from the \hho modeling (Section~\ref{sec:water}) as $M_{ISM}=\mu (4/3) \pi R^2 m_H N_H$, where $\mu=1.4$ accounts for the mass in helium, $R$ is the radius, $m_H$ is the mass of hydrogen.
For $R=500$ pc, we calculate $M_{ISM}=7.5\times10^{9}$ and an implied $M_{ISM}/M_{dyn}=0.47$.
This ISM mass estimate is sensitive to the best-fit dust temperature and $\tau_{100}$ inferred from the \hho modeling. Higher temperatures and lower $\tau_{100}$ values would further reduce the estimate ISM mass.

Using the starburst size, inferred $\Sigma_{IR,50}$, and gas fraction of $f_g\approx0.3$ (as observed in sources with similar \LIR), we can compare with the \citet{Thompson2005} model for a radiation pressure-limited starburst.
We find that \source is near to but slightly below the maximal starburst line for these values (for $\sigma\approx200$ \kms).

For the measured peak flux density of $12.6$ mJy beam$^{-1}$ the brightness temperature at $333$ GHz is $2.7$ K, suggesting the dust emission on the scale of our beam is optically thin.
If we take the $40$ K dust temperature from modeling of the \hho lines as the true dust temperature, we estimate $\tau=0.071$.

\subsection{Comparison to Planck Limit and Herschel Extrapolation}

\begin{figure}
\includegraphics[width=0.48\textwidth]{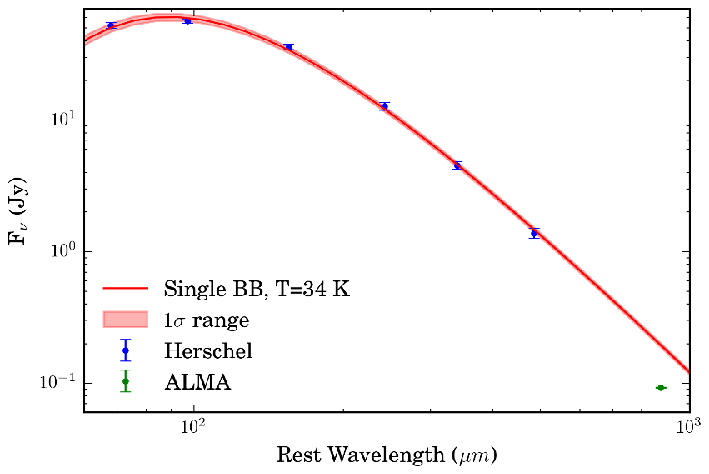}
\includegraphics[width=0.48\textwidth]{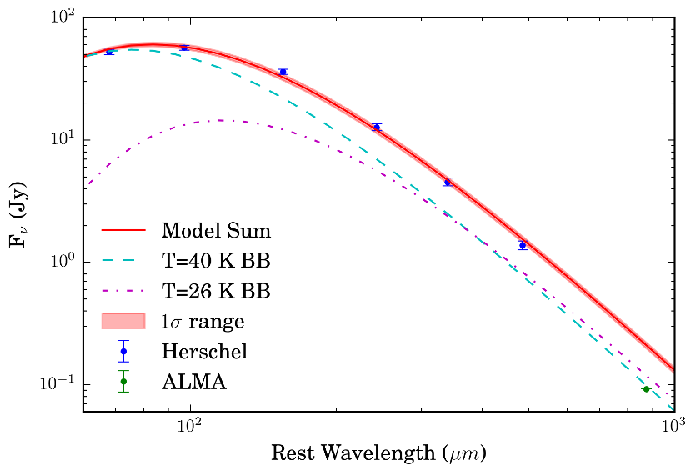}
\caption{Herschel PACS and SPIRE spectral energy distribution of \source (fluxes from J. Chu et~al. \emph{in preparation}), fit with two different models.
Top: Single-component modified blackbody fit with a best-fit temperature of $34$ K.
Bottom: Two-component modified blackbody fit. One component was fixed in temperature to 40 K, as derived from our \hho modeling and the amplitude fixed to that of the ALMA $333$ GHz flux measurement. The temperature and amplitude of the second component was left free and the sum was fit to the Herschel measurements. The best-fit temperature for the second component is $26$ K.
For all modified blackbodies we fix the dust emissivity to $\beta=1.8$.
Both fits to the Herschel photometry predict a similar $333$ GHz flux density, 190 mJy from the single-component and 210 mJy for the two-component model.
Both predictions are broadly consistent with the Planck upper limit \citep{PlanckXXVIII} and would suggest that these ALMA observations are resolving out approximately half the flux, which would be extended on scales $>4\arcsec$.}
\label{fig:SED}
\end{figure}

Our ALMA detection is consistent with the Planck $357$ GHz non-detection, which reported an upper limit of $\sim200$ mJy \citep{PlanckXXVIII}.
For an additional estimate of the total $333$ GHz continuum emission we fit the Herschel PACS and SPIRE photometry (J. Chu et~al. in preparation) with a single-component modified blackbody.
We fix the dust emissivity to $\beta=1.8$ \citep[e.g.,][]{Planck2011dust}.
and derive a best-fit temperature, $T_{BB}=33\pm0.5$ K\footnote{Errors obtained using MCMC exploration with the \texttt{emcee} package \citep{Foreman-Mackey2013} and represent the $1\sigma$ range.}.
Using this modified blackbody fit we predict a $333$ GHz flux of $\sim190$ mJy (Figure~\ref{fig:SED}, top).

If this extrapolation is correct, it suggests were are resolving out up to $\sim50\%$ of the 333 GHz flux with our ALMA observations.
This missing flux should be extended on scales $>4\arcsec$ ($>2.6$ kpc; the scale on which our observations will be affected by spatial filtering) and $<36.2\arcsec$ ($<23.7$ kpc; the SPIRE beam at $500~\mu m$), since we are using galaxy-integrated values.
\footnote{We note that \source is unresolved in all the Herschel PACS and SPIRE bands, including the $70~\mu m$ band, which has a resolution of $5.6\arcsec$ (3.7 kpc).}
This deficit of flux corresponds to an additional $M_{\textnormal{ISM}}\approx3.2\times10^{10}$ \Msun (following the \citealt{Scoville2014} empirical relation).
Taking $4\arcsec$ as a lower-limit on the scale for all of the missing flux, this implies the $\Sigma_{ISM}$ of any missing, extended component must be $<10^9$ \Msun kpc$^{-2}$.
This is at least a factor of 10 lower than the mass determined for the inner kpc, so our estimate of $\Sigma_{ISM, 1kpc}$ is unlikely to be strongly biased by missing continuum flux, and uncertainties our estimates of the ISM mass surface density are likely dominated by the scatter in the M$_{ISM}-L_{850~\mu m}$ relation.

Using the Herschel-derived temperature and the extrapolated flux, the dust mass would be $M_{dust,extrapolated}=8.2\times10^{8}$ \Msun.
Assuming a gas-to-dust ratio of 100 implies an ISM mass of $M_{ISM,extrapolated}=8\times10^{10}$ \Msun, a factor of 2.6 higher than inferred from our ALMA observations.

The dust temperature inferred from the Herschel photometry ($34$ K) is lower than that obtained from the modeling of \hho lines ($40$ K).
This is likely due to the fact that the \hho lines are tracing the denser molecular gas while the Herschel continuum measurements with a larger beam and fewer spatial filtering issues likely include emission from cooler, diffuse dust.

To further explore this, we also fit the Herschel measurements with a two-component model, consisting of a 40 K modified blackbody normalized to our measured ALMA flux and a second modified blackbody with the temperature and normalization left as free parameters.
For both components, we left $\beta$ fixed to 1.8.
This two-component model predicts a $333$ GHz flux of $\sim210$ mJy (Figure~\ref{fig:SED}, bottom), similar to that of the single-component model and roughly consistent with the Planck upper limit.

The best-fit temperature of the second blackbody component is $26$ K.
This is consistent with dust temperatures for normal galaxies \citep[e.g.,][]{Skibba2011}.
Because the 40 K blackbody was normalized to the integrated ALMA flux, the excess flux from the 26 K blackbody, if present and resolved out by our ALMA observations, should be extended on scales $>4\arcsec$.
This two-component model is thus consistent with a scenario in which (U)LIRGs have concentrated nuclear starbursts which are surrounded by less intense star formation that more closely resembles normal galaxies.
This is consistent with the results of \citet{Diaz-Santos2014}, who found the extended \cii emission of IR-selected galaxies to be similar to normal star forming galaxies, even if the nuclear starbursts show pronounced \cii deficits.
In the \citet{Diaz-Santos2014} sample, this extended star formation was detected on scales of $1-12.6$ kpc, and thus the scale of this diffuse star formation is comparable to the physical scale of any flux resolved out by our ALMA observations ($2.6$ kpc).

We note that the $40$ K blackbody, anchored to the ALMA data-point, falls below the observed $14~\mu m$ flux \citep{Inami2013}, meaning an additional higher-temperature component would be needed to fit the mid-infrared portion of the SED, though the same is true for the single-component $T=34$ K fit to the Herschel data.

We prefer the 2-component fit to the Herschel bands over the single-component fit, due to its more realistic representation of the likely physical origin of the thermal dust emission in \source.
The 40 K component dominates the IR luminosity of the system (Figure~\ref{fig:SED}, bottom), thus we conclude the $333$ GHz emission is a reasonable proxy for the distribution of the \emph{nuclear} starburst containing $\sim80\%$ of \LIR and that the size offers a reliable way to estimate $\Sigma_{IR}$.
Continuum observations at $\sim333$ GHz with a single-dish telescope or ALMA in a compact configuration would provide a reliable measurement of the total flux at this frequency and a definitive statement on how much flux is resolved out with our observations.
In addition, ALMA observations would afford a measure of distribution of any flux missing in these observations.

\section{The Dense Molecular Gas and the Nuclear Activity in \source}
\label{sec:activity}

The sub-mm continuum emission tracing the starburst has a diameter of $\sim500$ pc (FWHM from the Gaussian fit).
This area of the starburst is associated with an elevated \HCN{4}{3}/\HCO{4}{3} ratio, and there is tentative evidence for a dense molecular wind.
What do these observations imply about the molecular gas properties and the nuclear activity in \source?

\subsection{The Excitation of the Dense Molecular Gas}

Of the physical processes discussed in Section~\ref{sec:moloverview} which could plausibly lead to enhanced HCN/HCO$^+$ emission, radiative pumping seems unlikely due to the lack of detectable HCN $v_2=1f$ emission.
Previous studies of global HCN/HCO$^+$ ratios in IR galaxies have found limited evidence to support XDRs as the driver of enhanced HCN emission \citep{Costagliola2011,Privon2015} and source compactness (traced by \cii/\LFIR) is only weakly correlated with HCN enhancements \citep{Privon2015}.
While we cannot model the conditions of the dense molecular gas in \source with the limited set of lines presented here, exploratory LVG modeling has suggested enhanced HCN abundances may be required to explain the observed HCN/HCO$^+$ ratio over the nucleus.
Several recent studies have proposed mechanical heating can explain HCN enhancements \citep{Loenen2008,Kazandjian2012,Izumi2016}; can mechanical heating explain this abundance enhancement in \source?

We can use the measured star formation properties to estimate the supernova rate and the associated mechanical heating rate in the starburst of \source.
The star formation rate of $170$ \Msun yr$^{-1}$ implies a supernova rate of approximately 1.2 yr$^{-1}$ \citep[for a Salpeter IMF;][]{Mattila2001}, and an associated injection of mechanical energy via shocks.
This star formation rate is mostly contained within the region of elevated HCN/HCO$^+$.
We similarly expect the supernovae to be concentrated within this region and so associated with the elevated HCN emission.
Using 40\% of the supernova rate estimated above (to match the fraction of \LIR expected to be contained within the starburst FWHM) and the size of the starburst from the millimeter continuum emission ($d_{SB}=500$ pc; Section~\ref{sec:continuum}), and Equation 1 from \citet{Kazandjian2012}, we estimate the mechanical energy injection from supernovae to be $\Gamma_{mech} > 8\times10^{-21}$ erg s$^{-1}$ cm$^{-3}$, for a PDR volume filling factor of 0.1.
At this level of $\Gamma_{mech}$ and above, the \citet{Kazandjian2012} models of mechanical heating in molecular clouds show substantial abundance enhancements of HCN relative to HCO$^+$ (3--300x) and are sufficient to explain the ratios we see here, and are consistent with our exploratory LVG modeling.

Significant non-gravitational motion of the gas would also serve to increase the turbulence in the dense gas disk, which has $\sigma\gtrsim100$ \kms (Figure~\ref{fig:kinematics}).
We also note that the observed outflow/wind, can also result in mechanical heating \citep[e.g.,][]{Izumi2016}, so the molecular wind seen here and by \citet{Veilleux2013} may further increase the mechanical heating rate above that from the supernovae.

The majority of systems with substantial \cii deficits observed by \citet{Gonzalez-Alfonso2015} show evidence for strong OH $65~\mu m$ absorption; \source is an outlier in this respect, with a significant \cii deficit but no observed OH $65~\mu m$ absorption.
\citet{Gonzalez-Alfonso2015} argue this OH $65~\mu m$ absorption requires warm dust (T$>50$ K) and high column densities (N$_{H}\gtrsim6\times10^{23}$ \cmcol).
Thus, the lack of OH $65~\mu m$ absorption suggests the dust is cooler and/or the column densities are somewhat low, consistent with the non-detection of \hho lines above $400$ K, the modeling results for the \hho lines (Section~\ref{sec:water}), and the non-detection of the $v_2=1f$ \HCN{4}{3} line.
The dust temperature of $40$ K obtained from modeling of the \hho lines is consistent with the non-detection of OH $65~\mu m$ absorption.

The general role of infrared pumping in enhancing $v_2=0$ lines is still uncertain, but our non-detection of the $v_2=1f$ \HCN{4}{3} line and the non-detection of OH65 suggest that the starburst in \source is too cool and/or has insufficient column density for HCN to absorb significant numbers of IR photons.
Though not conclusive, this suggests that radiative pumping is not a significant contributor to the enhanced HCN emission.

These observations and associated studies of \source are consistent with a scenario in which the molecular gas is primarily experiencing collisional excitation with the enhanced HCN emission originating from an elevated abundance caused by mechanical heating from the ongoing starburst.
Confirmation of the abundance enhancement in the nucleus of \source will require new multi-transition HCN and HCO$^+$ maps to facilitate more detailed LVG modeling.

In contrast to the nuclear regions, the extended regions ($\gtrsim300$ pc) of the molecular emission have HCN/HCO$^+\approx1$.
Thus, it appears that the more extended molecular emission is consistent with typical PDR environments in star forming galaxies.

\subsection{The Origin and Fate of the Dense Molecular Wind}

Will this dense outflow escape the galaxy or is it doomed to return to the nucleus?
Using the dynamical mass obtained in Section~\ref{sec:kinematics} for the inner kpc, the escape velocity at 1 kpc is $(GM/R)^{0.5}\sim600$ \kms.
The outflow velocities we see in the dense molecular gas reach a maximum of $\sim300$ \kms (Figure~\ref{fig:outflow}).
Thus, it appears that these winds will be unable to escape the nucleus, unless the outflow axis is highly inclined along the line of sight ($\theta\lesssim35^{\circ}$), reducing the projected radial velocity component.
This observed wind, or at least its dense molecular component, appears destined to stall and remain bound to the system.

Dense gas conversion factors have been determined for ``normal'' galaxies, under the assumption that the emission is coming from an ensemble of distinct clouds \citep{Gao2004}.
Estimating the mass outflow rate of dense material in \source is problematic given the uncertain excitation and abundance of these dense gas tracers, as discussed above.
Additionally, the ensemble of clouds assumption may not be appropriate for the nuclear ISM in ULIRGs, where the molecular gas may have a more smooth distribution \citep[e.g.,][]{Solomon1997,Scoville2014}.

However, we can provide an \emph{upper limit} to the dense mass outflow rate, by using the ensemble of clouds approximation.
If we assume the emission is optically thick and thermalized up to the 4--3 line, the equivalent 1--0 luminosity of the outflow will equal that of the 4--3 line: \Lprime{HCN}{1}{0}=\Lprime{HCN}{4}{3}=$2\times10^{7}$ K km s$^{-1}$ pc$^2$.
This would correspond to a dense molecular mass of $2\times10^{8}$ \Msun in the outflow.

Taking this upper limit on the outflowing dense gas mass, we can compute the mass outflow rate, following Equation~1 of \citet{Cicone2014} and using the measured outflow extent of $r=0.2$ kpc and outflow velocity of $250$ \kms.
We obtain an \emph{upper limit} on the mass of outflowing \emph{dense} molecular gas of $\dot{M}_{dense}<770$ \Msun yr$^{-1}$, significantly in excess of the $170$ \Msun \pyr SFR.
Because the outflow appears to follow the rotation pattern (but lie above the flattening of the rotation curve), the velocity of the material identified as outflowing may have an important rotational component to its velocity, which would reduce the inferred $\dot{M}$.

We emphasize here that the above calculations merely represent an upper limit to the mass of dense gas detected here and the actual mass could be at least order of magnitude lower \citep[e.g.,][]{Aalto2015}.
That the outflowing gas shows a high HCN/HCO$^+$ ratio ($>3$) is strong evidence that the excitation and/or abundance is non-standard and that galactic conversion factors are unlikely to be accurate.
An example of this is Mrk~231, where the HCN, HCO$^+$, and HNC species do not appear to be co-spatial, but instead reflect an outflow with significant abundance variations in the wind \citep{Lindberg2016}, suggesting significant chemical effects which need to be considered for the appropriate interpretation of these molecules in outflows.

Within the substantial uncertainties in our mass outflow rate and the important consideration that we are only diagnosing the dense component of the outflow, the observations are broadly consistent with either an AGN or starburst driven wind.
CO observations (K.~Sliwa \emph{in preparation}) will be needed to both confirm the molecular outflow, estimate the total molecular outflow rate, and the dense fraction.

The outflow kinematics, relative to the disk may shed some light on the origin of the outflow.
A starburst-driven outflow would be expected to exit perpendicular to the disk \citep[e.g.,][]{GarciaBurillo2001,Walter2002,Bolatto2013b}, as the path of least resistance.
In this case, there should be a $\sim90^{\circ}$ shift in the PA of the disk, compared to the PA of the outflow.
The HCN line wings in \source have a similar PA to that of the disk (Figure~\ref{fig:outflow}) suggesting the wind is not exiting perpendicular to the disk.
However, AGN outflows may be more collimated by the small-scale accretion disk and/or torus.
This small-scale collimation may have a random orientation with respect to the kpc-scale disk, so an AGN-driven outflow may be a more natural explanation for the wind identified here.

\source was detected in OH $119~\mu m$ absorption using Herschel observations \citep{Veilleux2013}, with maximum outflow velocities of up to 1200 \kms.
The OH outflow is extends to significantly higher velocities than seen in the ALMA data.
It is unclear if the outflow in this system is similar to the multi-phase outflow seen in Mrk~231 \citep{Aalto2015}, where the HCN outflow corresponds to the lower-velocity OH outflow component, while the higher-velocity OH outflow does not have a corresponding HCN outflow.
The velocity difference between OH and HCN outflow is broadly consistent with a fast, diffuse outflow (seen in OH) that is entraining dense clouds at lower velocity.

\HCN{4}{3} emission associated with the lower-velocity portion of a CO-detected molecular outflow has been detected in Mrk~231 \citep{Aalto2012,Aalto2015}, NGC~1068 \citep{Garcia-Burillo2014}, and M~51 \citep{Matsushita2015}.
The outflows are thought to be jet-driven (M~51) or AGN-driven (Mrk~231, NGC~1068) based on the energetics or the inferred mass outflow rates.
In these cases and in \source, the material identified as outflowing has a high HCN/HCO$^+$ ratio and is consistent with expectations from mechanical heating \citep[e.g.,][]{Matsushita2015,Lindberg2016}.

\subsection{The Relationship Between $\Sigma_{IR}$ and $\Sigma_{14}$}

Based on the sizes measured from the sub-mm continuum, we have inferred a high $\Sigma_{IR}$, indicating the presence of a fairly extreme starburst.
In contrast, computing $\Sigma_{14}$ using the same size, we find a somewhat unremarkable luminosity surface density, relative to sources which feature IR pumping of HCN.

\begin{figure}
\includegraphics[width=0.45\textwidth]{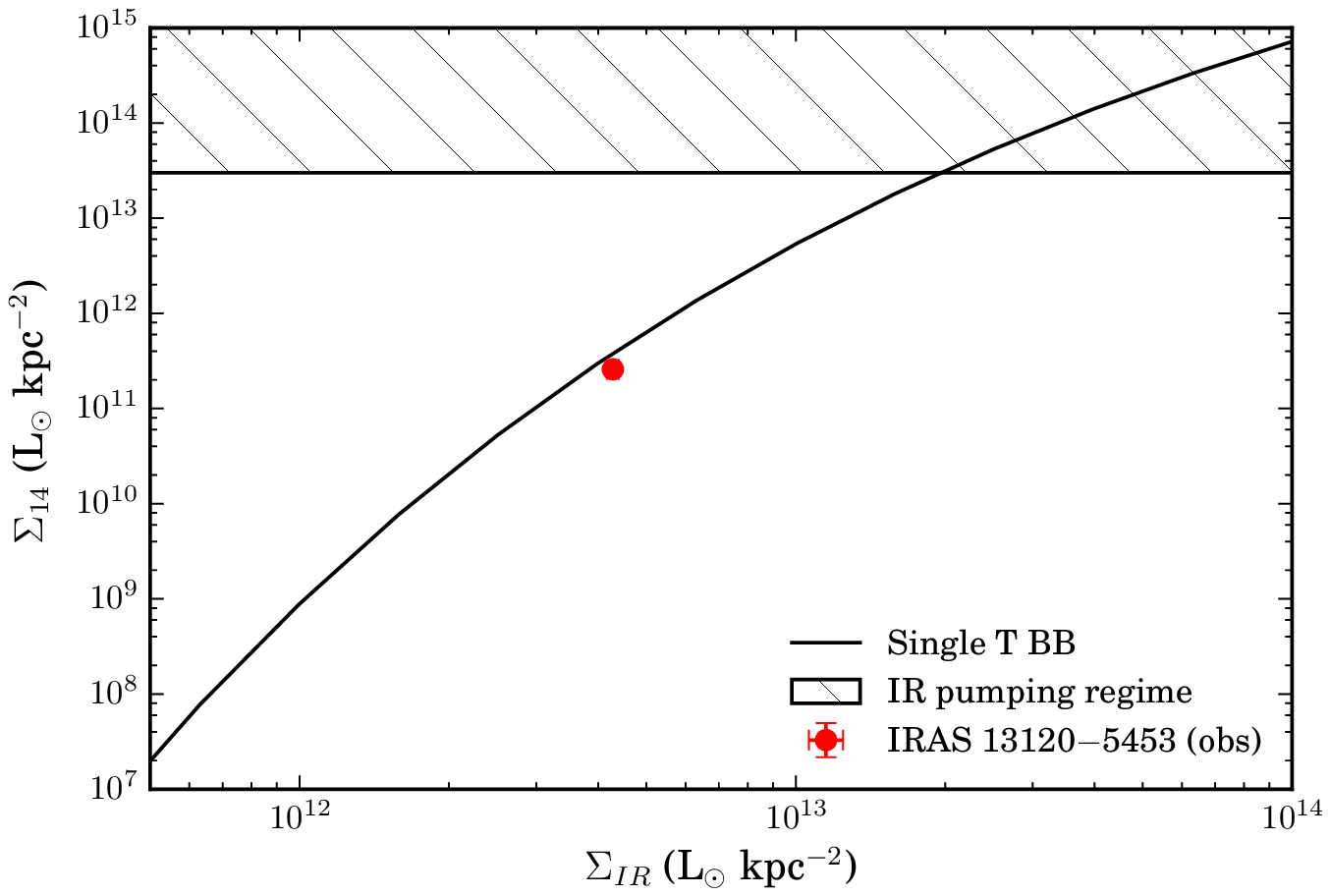}
\includegraphics[width=0.45\textwidth]{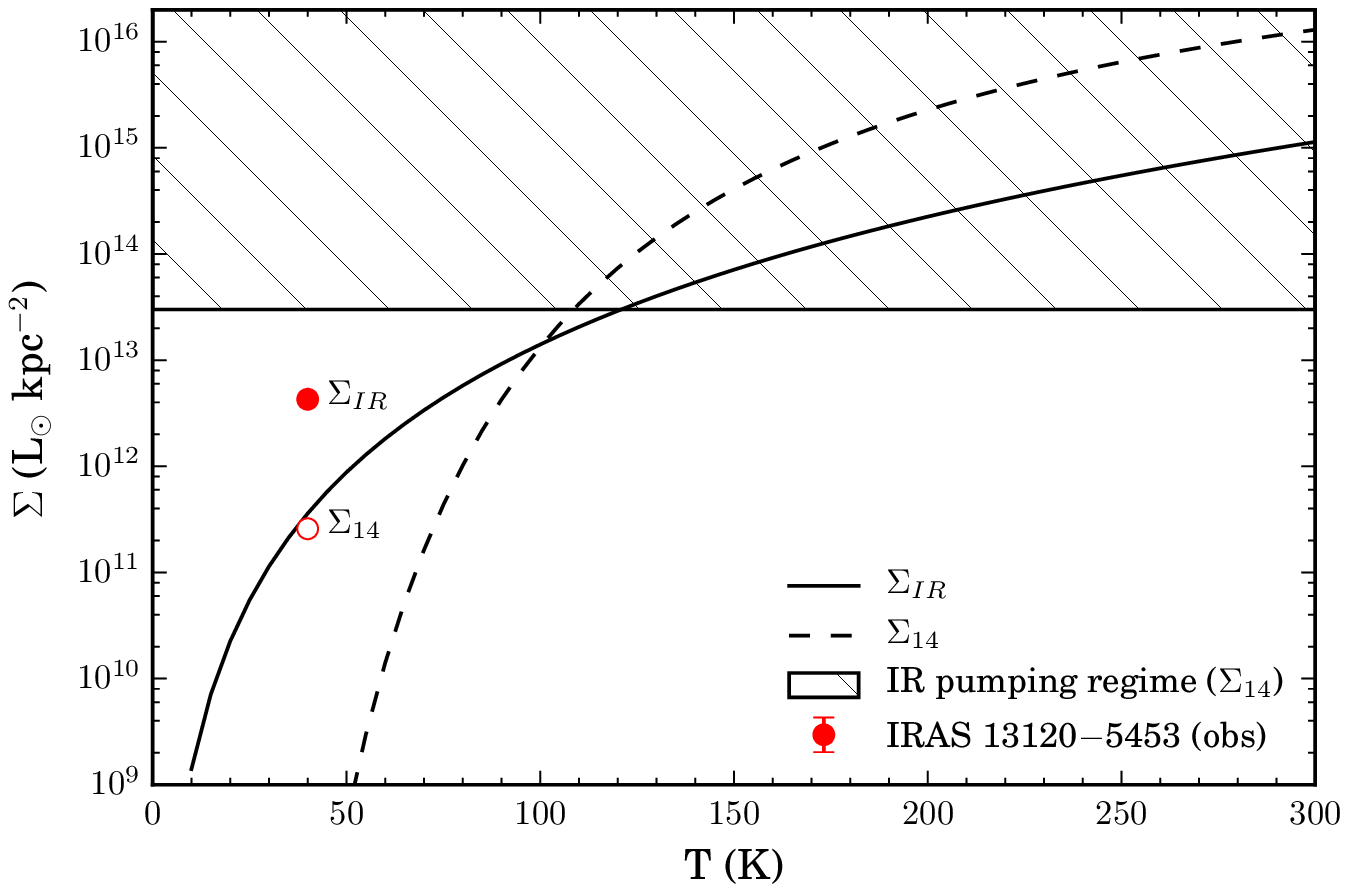}
\caption{Top: The relationship between $\Sigma_{IR}$ and $\Sigma_{14}$ for a single component blackbody.
The temperature increases as a function of $\Sigma_{IR}$ and $\Sigma_{14}$.
The red point marks the observed $\Sigma_{IR}$ and $\Sigma_{14}$ values for \source. The error bars are smaller than the point size.
Bottom: Explicit dependence of $\Sigma_{IR}$ and $\Sigma_{14}$ on T for a single component blackbody.
For reference, $B_{\nu}(\nu, T)$ peaks at $\lambda_{peak}=14~\mu m$ (21.4 THz) when T$=365$ K.
In both figures, the hatched region marks where $\Sigma_{14}$ is high enough to potentially drive radiative pumping of HCN \citep[note that pumping also requires sufficiently high column densities;][]{Aalto2015a}.
For T $=40-100$ K, $\Sigma_{IR}$ could be consistent with radiation pressure limited starbursts \citep[modulo gas fraction and velocity dispersion;][]{Thompson2005} without $\Sigma_{14}$ being high enough to support radiative pumping of HCN.
The observed $\Sigma$ values for \source (bottom panel) suggests the single temperature model is not a complete descriptor of the infrared-emitting region.
On these scales, the errorbars for the \source values are smaller than the point sizes.
}
\label{fig:sigmas}
\end{figure}

Some insight into the apparent discrepancy can be gained by examining the $\Sigma_{IR}$ and $\Sigma_{14}$ for an idealized single-component blackbody.
In particular both quantities have different dependence on the temperature: $\Sigma_{IR}\propto T^4$, and $\Sigma_{14}\propto(e^{h(21.4~\mathrm{THz})/(k_b T)} - 1)^{-1}$.
In Figure~\ref{fig:sigmas} we show the relationship between $\Sigma_{IR}$ and $\Sigma_{14}$ for a single-component blackbody, and the explicit dependence of these luminosity surface densities on the temperature.
In the upper panel we plot the observed values for \source and shade the range of $\Sigma_{14}$ values which may result in IR pumping of HCN \citep[in the presence of sufficiently high column densities;][]{Aalto2015a}.
Note that, in the top panel, there is an implicit dependence of $\Sigma_{IR}$ and $\Sigma_{14}$ on T.
While we do not explicitly plot $\Sigma_{13}$ (potentially relevant for the as-yet undetected IR pumping of HCO$^+$) the curve closely tracks that of $\Sigma_{14}$.

These figures demonstrate that, for even fairly high $\Sigma_{IR}$ systems which could be radiation pressure limited starbursts \citep[i.e., those between $10^{12}-10^{13}$ \Lsun kpc$^{-2}$, with the exact value depending on $\sigma$ and $f_{gas}$;][]{Thompson2005}, $\Sigma_{14}$ can remain below the range required for radiative pumping of HCN.
Thus, some systems with radiation pressure limited starbursts may not show vibrational HCN emission, unless they have an embedded compact core with $\Sigma_{14}>3\times10^{13}$ (corresponding to $\Sigma_{IR}\gtrsim2\times10^{13}$ \Lsun kpc$^{-2}$ for a single-component blackbody).

\subsection{A Scenario for the Nuclear Activity in \source}

The overall properties of \source are nearly consistent with the properties of these CONs, notably the presence of a Compton-thick AGN \citep{Teng2015} and a significant \cii deficit \citep{Diaz-Santos2014}.
However, an important difference between \source and the CONs is the lack of vibrational HCN emission and a substantially lower estimated $14~\mu m$ luminosity surface density.
Despite the high HCN/HCO$^+$ ratio and the \cii deficit, \source does not seem to be as extreme as the systems presented by \citet{Aalto2015a}.
We note that while the $40$ K dust temperature is higher than typical nearby galaxies, which have $T_{dust}\sim25$ K \citep[e.g.,][]{Skibba2011}, this is similar to dust temperatures in local IR-selected galaxies (R. Herrero-Illana \emph{in preparation}) and much cooler than is seen in CONs such as Arp~220, NGC~4418, Mrk 231, and Zw 049.057 \citep[T$_{dust}=90-130, 130-150, 110,$ and $>100$ K, respectively;][]{Gonzalez-Alfonso2012,Gonzalez-Alfonso2013,Gonzalez-Alfonso2014,Falstad2015}.

The elevated \HCN{4}{3}/\HCO{4}{3} ratio is consistent with a HCN abundance enhancement from mechanical heating, implying a turbulent nuclear ISM.
The outflows in this system suggest the feedback from star formation and/or the AGN is pushing the nuclear ISM to larger radii.
One explanation for the lack of $v_2=1f$ emission is that this feedback has sufficiently inflated the nuclear gas concentration to the point where the $14~\mu m$ luminosity density is too low to drive IR pumping.
Stated another way, the movement of gas (and associated dust) to larger distance from the central starburst may effectively reduce the dust temperature by reducing the optical depth to UV and IR photons.

If the CON phase is a universal (but likely short-lived) phase in a luminous merger, is \source in a pre- or post-CON state?
Given the advanced dynamical stage of the system \citep[post-merger;][]{Haan2011,Stierwalt2013} and the presence of loops, it is likely the nucleus will continue to receive a modest inflow of gas as the tidal material continues to re-accrete \citep[e.g., as seen in NGC~7252;][]{Hibbard1995}.
However, it seems unlikely that the galaxy will experience substantial large-scale tidal torques which would drive significant additional quantities of gas into the nucleus on a short timescale, such as what likely triggered the present starburst \citep[e.g.,][]{Barnes1991}.
That the starburst appears compact, yet lacks significant HCN $v_2=1f$ emission, has a $\Sigma_{14~\mu m}$ lower than CONs, and shows evidence for outflows, suggests (if \source experienced at CON phase) we may be seeing \source in a post-CON phase where feedback is clearing the nucleus.
This is in qualitative agreement with the results of \citet{Aalto2015a}, finding that systems with the most luminous $v_2=1f$ lines (relative to \LIR) lack outflows, and suggests a CON phase precedes the onset of significant outflows.
If the feedback is starburst-driven, this may imply future starburst episodes as the dense gas wind stalls and returns \citep[e.g.,][]{Torrey2016}.

One remaining question regarding the outflow concerns the $\sim1200$ \kms outflow seen in OH \citep{Veilleux2013}: does this fast component of the outflow contain a significant amount of mass which will become unbound from the system?
If the OH outflow represents only the tail of the distribution of outflowing material, the bulk of the ISM will remain bound.
What fraction of the ISM is escaping will impact whether the feedback is regulating or quenching star formation.
Given $\sim30\%$ of the dynamical mass in the central kpc is currently gas, determining if that gas is permanently removed from the system has significant implications for the structure of the remnant galaxy and the growth of the central SMBH.

\section{Conclusions}

We present ALMA Band 7 observations of \HCN{4}{3} ($v=0$ and $v_2=1f$), \HCO{4}{3}, \CS{7}{6}, and the 333 GHz continuum, and Herschel observations of \hho lines.
We do not detect the $v_2=1f$ line of HCN and the ratio of the $v=0/v_2=1f$ lines is $>10\times$ higher than seen in sources with detectable IR pumping of HCN.
Using these data, we show:

\begin{itemize}
  \item{The HCN/HCO$^+$ ratio is elevated over the nucleus and is consistent with the expectations of increased abundance due to mechanical heating by supernovae, though the AGN and the outflows may contribute additional turbulent energy.}
  \item{Line wings on the HCN and HCO$^+$ profiles provide evidence for the presence of a dense molecular outflow which exceeds the virial range for the gas disk but will remain bound to the system.
  The large uncertainties in the abundance and excitation of HCN hamper estimates of the mass and means we cannot distinguish between an AGN or starburst origin based on energetics.}
  \item{Our modeling of the \hho emission suggests a nuclear ISM that has a moderate dust temperature (40 K), higher than normal galaxies but a factor of $>2$ cooler than T$_{dust}$ inferred for CONs.}
  \item{We marginally resolve the 333 GHz continuum emission, which implies a starburst diameter of $\sim500$ pc and an infrared surface density of $\Sigma_{IR}=4.7\times10^{12}$ \Lsun kpc$^{-2}$.
  This is near to, but below predictions from radiation pressure limited starburst models.}
  \item{A potentially large fraction of the dynamical mass within the central kpc is gaseous, making the fate of this nuclear gas critical to predicting the long-term evolution of the structure of \source and the growth of its central SMBH.}
\end{itemize}

Combining this evidence, we propose \source is being observed during a period of nuclear activity where feedback is inflating the nuclear gas configuration and restricting star formation.
The dense molecular outflow will likely not escape the system and so will return to the nucleus to fuel a future episode of star formation and SMBH accretion.
The observed HCN/HCO$^+$ ratio is consistent with mechanical energy injection from supernovae, though the AGN and outflows may also contribute.

\acknowledgements

The authors thank the anonymous referee for his/her comments, which have improved the quality of the paper.
This paper makes use of the following ALMA data: ADS/JAO.ALMA\#2012.1.00817.S.
ALMA is a partnership of ESO (representing its member states), NSF (USA) and NINS (Japan), together with NRC (Canada) and NSC and ASIAA (Taiwan) and KASI (Republic of Korea), in cooperation with the Republic of Chile.
The Joint ALMA Observatory is operated by ESO, AUI/NRAO and NAOJ.
The National Radio Astronomy Observatory is a facility of the National Science Foundation operated under cooperative agreement by Associated Universities, Inc.

Herschel is an ESA space observatory with science instruments provided by European-led Principal Investigator consortia and with important participation from NASA.

G.C.P. was supported by a FONDECYT Postdoctoral Fellowship (No.\ 3150361) and acknowledges the hospitality of the National Socio-environmental Synthesis Center (SESYNC), where portions of this manuscript were written.
G.C.P. and E.T. acknowledge support from the CONICYT Anillo project ACT1101.
F.C. acknowledges support from Swedish National Science Council grant 637-2013-7261.
S.G.B. acknowledges support from Spanish grants AYA2013-42227-P and ESP2015-68964-P.
K.S. acknowledges grant 105-2119-M-001-036 from Taiwanese Ministry of Science and Technology.

Portions of this work were performed at the Aspen Center for Physics, which is supported by National Science Foundation grant PHY-1066293.
This work was partially supported by a grant from the Simons Foundation.

The authors thank Claudia Cicone, Francois Schweizer, and Claudio Ricci for comments on an earlier version of the manuscript.
G.C.P. thanks Paul Torrey, David Patton, Chris Hayward, Desika Narayanan, Claudia Cicone, Nick Scoville, and Patricia Bessiere for helpful discussions.

This research has made use of the NASA/IPAC Extragalactic Database (NED) which is operated by the Jet Propulsion Laboratory, California Institute of Technology, under contract with the National Aeronautics and Space Administration.
This research has made use of NASA's Astrophysics Data System.
This research made use of ipython \citep{Perez2007}, numpy \citep{Vanderwalt2011}, Astropy \citep[\url{http://www.astropy.org}, a community-developed core Python package for Astronomy]{astropy}, \texttt{emcee} \citep{Foreman-Mackey2013}, and the \texttt{dust\_emissivity} package (\url{https://github.com/keflavich/dust_emissivity}).
The figures in this paper were created using matplotlib \citep{Hunter2007} and APLpy \citep[an open-source plotting package for Python hosted at \url{http://aplpy.github.com}]{aplpy}.

\bibliography{ms}

\end{document}